\documentclass[aps,pre,reprint,superscriptaddress]{revtex4-2}

\usepackage[utf8]{inputenc}

\usepackage{amsmath}
\usepackage{amssymb}
\usepackage{bm}
\renewcommand{\vec}{\bm}

\usepackage{hyperref}

\usepackage{graphicx}

\begin{document}


\title{Laminar chaos in systems with quasiperiodic delay}


\author{David Müller-Bender}
\email[]{david.mueller-bender@mailbox.org}
\affiliation{Institute of Physics, Chemnitz University of Technology, 09107 Chemnitz, Germany}
\author{Günter Radons}
\email[]{radons@physik.tu-chemnitz.de}
\affiliation{Institute of Physics, Chemnitz University of Technology, 09107 Chemnitz, Germany}
\affiliation{ICM - Institute for Mechanical and Industrial Engineering, 09117 Chemnitz, Germany}


\date{\today}

\begin{abstract}
A new type of chaos called laminar chaos was found in singularly perturbed dynamical systems with periodic time-varying delay [Phys. Rev. Lett. 120, 084102 (2018)].
It is characterized by nearly constant laminar phases, which are periodically interrupted by irregular bursts, where the intensity levels of the laminar phases vary chaotically from phase to phase.
In this paper, we demonstrate that laminar chaos can also be observed in systems with quasiperiodic delay, where we generalize the concept of conservative and dissipative delays to such systems.
It turns out that the durations of the laminar phases vary quasiperiodically and follow the dynamics of a torus map in contrast to the periodic variation observed for periodic delay.
Theoretical and numerical results indicate that introducing a quasiperiodic delay modulation into a time-delay system can lead to a giant reduction of the dimension of the chaotic attractors.
By varying the mean delay and keeping other parameters fixed, we found that the Kaplan-Yorke dimension is modulated quasiperiodically over several orders of magnitudes, where the dynamics switches quasiperiodically between different types of high- and low-dimensional types of chaos.
\end{abstract}


\maketitle


\section{Introduction}

Processes that involve transport or evolution by a finite velocity are characterized by time-delays.
Time-delay systems are widely applied to model such processes and appear in many areas of science \cite{kuang_delay_1993,scholl_handbook_2007,erneux_applied_2009,lakshmanan_dynamics_2011,scholl_control_2016} and engineering \cite{erneux_applied_2009,stepan_retarded_1989,michiels_stability_2007}.
Beyond the well established mathematical theory  \cite{hale_introduction_1993,diekmann_delay_1995,hale_dynamics_2002}, an overview of recent advances of the theory and applications of time-delay systems can be found in the theme issues introduced by \cite{just_delayed_2010,erneux_introduction_2017,otto_nonlinear_2019}.
A review on chaos in time-delay systems can be found in \cite{wernecke_chaos_2019}.
Since the delay generating processes are in general influenced by environmental fluctuations or by the state of the considered system itself, the delays are in principle time- and state-dependent.
Generalizing the widely studied case of a constant delay and simplifying the challenging case of a state-dependent delay, a time-dependent delay can be considered, which is realistic if the delay generating process is nearly independent of the state of the system.
It is known that introducing a temporal delay variation increases the complexity of time-delay systems \cite{radons_complex_2009,lazarus_dynamics_2016}, which can improve the security of chaos communication \cite{kye_characteristics_2004,ghosh_synchronization_2007,kye_information_2012}.
A delay variation can induce different types of synchronization \cite{kye_synchronization_2004,kye_synchronization_2004_2,ambika_anticipatory_2009,ghosh_generalized_2009,ghosh_projective-dual_2011,senthilkumar_delay_2007,khatun_synchronization_2022}, it can stabilize \cite{madruga_effect_2001,otto_stability_2011,otto_application_2013} and destabilize systems \cite{louisell_delay_2001,papachristodoulou_stability_2007}, and influences mathematical properties such as the analyticity of solutions \cite{mallet-paret_analyticity_2014}.
Also effects on delayed feedback control \cite{gjurchinovski_stabilization_2008,gjurchinovski_variable-delay_2010,jungling_experimental_2012,gjurchinovski_delayed_2013} and on amplitude death in oscillator networks \cite{gjurchinovski_amplitude_2014} were studied.
While fast time-varying delays can be approximated by constant distributed delays \cite{michiels_stabilization_2005} and the stability of systems with slowly time-varying delays can be derived from the stability of constant delay systems \cite{otto_application_2013}, the intermediate case induces features that are not known from constant delay systems.
In \cite{otto_universal_2017,muller_dynamical_2017} it was demonstrated that there are basically two types of periodically time-varying delays, where one of the types leads to large differences from the known behavior of constant delay systems in the tangent space dynamics such as the scaling of the Lyapunov spectrum and the structure of the Lyapunov vectors.
Moreover in \cite{muller_laminar_2018,muller-bender_resonant_2019} it was shown that this delay type leads to a previously unknown type of chaotic dynamics called \emph{laminar chaos}, which is characterized by nearly constant laminar phases, whose intensity levels vary chaotically from phase to phase and the phases are periodically interrupted by short irregular bursts.
This comparably low-dimensional behavior differs drastically from the high-dimensional chaotic dynamics observed in the same systems for constant delay, which is characterized by high-frequency oscillations.
The first experimental observation of laminar chaos in an optoelectronic system \cite{hart_laminar_2019,muller-bender_laminar_2020}, where its robustness against noise was demonstrated, was followed by further experimental observations in electronic systems \cite{jungling_laminar_2020,kulminskii_laminar_2020}.
The synchronization of laminar chaotic systems was investigated in \cite{khatun_synchronization_2022}, and in \cite{kulminskiy_laminar_2022}, for the first time, laminar chaos was found in a constant delay system that is coupled to a laminar chaotic time-varying delay system.
In this paper, we generalize the theory on laminar chaos to systems with quasiperiodically time-varying delay.
Such delays are relevant, for instance, in the analysis of quasiperiodic solutions of systems with state-dependent delay \cite{he_construction_2017,he_construction_2016} or can be viewed as an intermediate step to understand systems with randomly time-varying delay, which are common in many systems \cite{verriest_stability_2009,krapivsky_stochastic_2011,gomez_stability_2016,qin_stability_2017,liu_stability_2019}.

We consider systems defined by the scalar delay differential equation
\begin{equation}
	\label{eq:sys}
	\frac{1}{\Theta} \dot{z}(t) + z(t) = f(z\bm{(}R(t)\bm{)}), \quad \text{with } R(t) = t-\tau(t),
\end{equation}
where $\tau(t)$ is the time-varying delay.
Systems with this structure and various nonlinearities $f(z)$ of the delayed feedback appear in many applications:
The Mackey-Glass equation \cite{mackey_oscillation_1977}, which is a model for blood-production, is given by $f(z)=\mu\,z/(1+z^{10})$, a sinusoidal nonlinearity, $f(z)=\mu\,\sin(z)$, gives the Ikeda equation \cite{ikeda_multiple-valued_1979, ikeda_optical_1980}, which first appeared as a model for light dynamics in a ring cavity with a nonlinear optical medium and also well describes certain optoelectronic oscillators \cite{hart_laminar_2019,larger_complexity_2013,chembo_optoelectronic_2019}, and the quadratic nonlinearity $f(z)=\mu\,z(1-z)$ was used to analyze general properties of such types of systems \cite{adhikari_periodic_2008}.
Chaotic diffusion can be observed with the climbing-sine nonlinearity $f(z)=z + \mu\,\sin(2\pi\, z)$ \cite{albers_chaotic_2022,albers_antipersistent_2022}.
Studies with further nonlinearities can be found in \cite{lakshmanan_dynamics_2011}.
If the parameter $\Theta$ is large, as we assumed in this paper, Eq.~\eqref{eq:sys} belongs to the class of singularly perturbed delay differential equations and large delay systems, which both are widely studied for constant delay \cite{ikeda_successive_1982,chow_singularly_1983,mallet-paret_global_1986,ikeda_high-dimensional_1987,mensour_chaos_1998,adhikari_periodic_2008,amil_organization_2015,wolfrum_eckhaus_2006,wolfrum_complex_2010,lichtner_spectrum_2011,giacomelli_coarsening_2012,marino_front_2014,faggian_evidence_2018,marino_excitable_2019,albers_antipersistent_2022} and also results on state-dependent delay are available \cite{mallet-paret_boundary_1992,mallet-paret_boundary_1996,mallet-paret_boundary_2003,kashchenko_local_2015,martinez-llinas_dynamical_2015}.
The separation into a large delay timescale and a small internal timescale plays a crucial role in the spatio-temporal representation of time-delay systems, which enables the observation of spatio-temporal phenomena in time-delay systems \cite{giacomelli_relationship_1996,yanchuk_spatio-temporal_2017,marino_spatiotemporal_2018,marino_spatiotemporal_2020}.
Using the concept of singularly perturbed time-delay systems, potentially high-dimensional systems can be easily implemented by opto-electronic systems, which is interesting for applications such as chaos communication \cite{goedgebuer_optical_1998,vanwiggeren_optical_1998,udaltsov_communicating_2001,keuninckx_encryption_2017}, random number generation \cite{uchida_fast_2008,reidler_ultrahigh-speed_2009,kanter_optical_2010}, and reservoir computing \cite{appeltant_information_2011,larger_high-speed_2017,hart_delayed_2019,stelzer_performance_2020}.
According to the concept of strong and weak chaos introduced in \cite{heiligenthal_strong_2011}, chaotic dynamics generated by Eq.~\eqref{eq:sys}, including the types of chaos considered here, can be classified as weak chaos since the linear instantaneous term $z(t)$ on the left hand side leads to a negative instantaneous Lyapunov exponent.

\begin{figure}
	\includegraphics[width=0.48\textwidth]{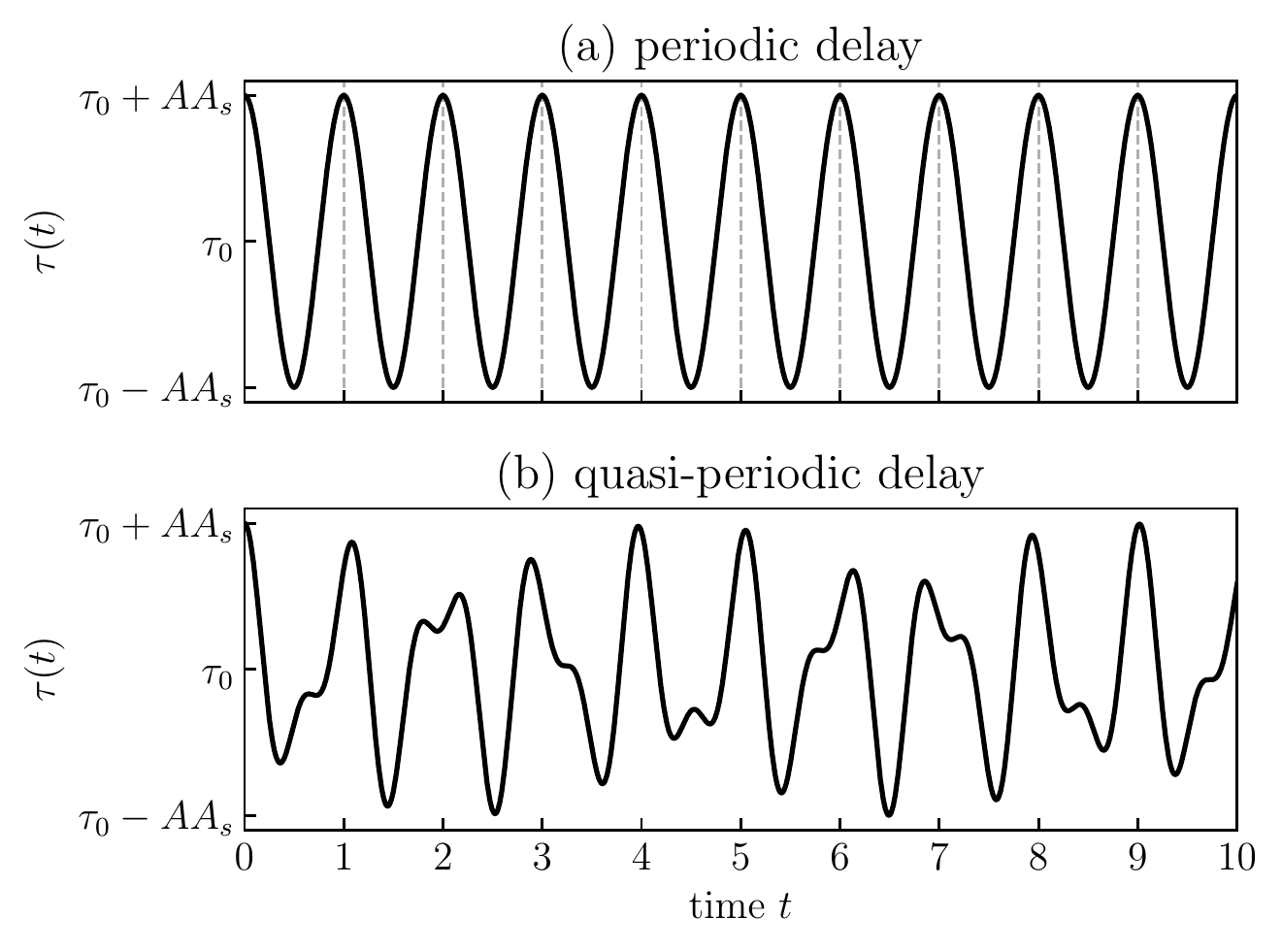}
	\caption{Exemplary time-varying delays according to Eq.~\eqref{eq:delay_def}.
		(a) Periodic delay ($N=1$) with frequency $\nu_1=1$ and (b) quasi-periodic delay with $N=2$ incommensurate frequencies $\nu_1=1$ and $\nu_2=\sqrt{\pi}$.
		The delays are parameterized by the mean delay $\tau_0$ and the amplitude parameter $A \in [0,1]$, which determines the largest bound $A\,A_s$ of the delay variation, where we have $A_s = (1/N) \sum_{n=1}^N (2\pi\nu_n)^{-1}$ (see Eq.~\eqref{eq:delay_def}).
	}
	\label{fig:delay}
\end{figure}

To generalize the theory on laminar chaos, we consider delays that are quasiperiodic in the sense of being an almost periodic function (cf. \cite{bohr_almost_1947}) instead of the periodic delay considered in \cite{muller_laminar_2018}.
Such type of delays can be defined by
\begin{equation}
	\tau(t) = \tau_0 + A\,g(\nu_1\,t,\nu_2\,t,\dots,\nu_N\,t),
\end{equation}
where $g : \mathbb{R}^N \to \mathbb{R}$ is $1$-periodic in all arguments and $N$ is the number of frequencies.
As in the original theory on laminar chaos, we assume that $\dot{\tau}(t)<1$ for almost all $t$, which avoids several mathematical problems and often can be motivated by physical arguments \cite{verriest_inconsistencies_2011,verriest_state_2012}.
For instance, in systems, where the delay is caused by a transport process, this assumption means that the distance a signal has to travel does not change faster than the velocity of the signal. 
The function $g$ must fulfill
\begin{equation}
	\label{eq:delayinvcond}
	1 > \dot{\tau}(t) = \sum_{n=1}^{N} \nu_n g_{(n)}(\nu_1\,t,\nu_2\,t,\dots,\nu_N\,t)
\end{equation}
for almost all $t$, where $g_{(n)}$ is the partial derivative of $g$ with respect to the $n$th argument.
For our numerical investigations, we consider delays of the form
\begin{equation}
	\label{eq:delay_def}
	\tau(t) = \tau_0 + \frac{A}{N} \sum_{n=1}^N \frac{\cos(2\pi\nu_n\,t)}{2\pi\nu_n}
\end{equation}
with the mean delay $\tau_0$ and the amplitude parameter $A$, where Eq.~\eqref{eq:delayinvcond} is fulfilled for almost all $t$ if we assume $A \in [0,1]$.
An exemplary periodic and a quasiperiodic delay generated by Eq.~\eqref{eq:delay_def} is shown in Fig.~\ref{fig:delay}(a) and (b), respectively.

The paper is structured as follows.
In Sec.~\ref{sec:review} the theory of laminar chaos is reviewed.
First numerical experiments with a quasiperiodic delay variation in Eq.~\eqref{eq:sys} are documented in Sec.~\ref{sec:lamchaos_quasi_num}, where it is demonstrated that laminar chaos exists in such systems.
In Sec.~\ref{sec:lamchaos_quasi_theory}, the theory of laminar chaos is generalized to systems with quasiperiodic delay by a rigorous analysis of the limiting system obtained for $\Theta=\infty$, which is needed to understand the numerical results, especially the differences to systems with periodically time-varying delay.
The effective dimension of the chaotic dynamics resulting from Eq.~\eqref{eq:sys} with quasiperiodic delay is considered in Sec.~\ref{sec:kydim}, where the Kaplan-Yorke dimension \cite{kaplan_chaotic_1979,farmer_dimension_1983} is numerically computed as a function of the mean delay $\tau_0$.
The differences of the structure in parameter space compared to periodically time-varying delay systems are elaborated and an outlook to systems with random delay is given.

\section{Review of laminar chaos}
\label{sec:review}

First we shortly recall the theory of laminar chaos from \cite{muller_laminar_2018,muller-bender_resonant_2019}.
In principle, our system of interest is a feedback loop, where a signal $z(t)$ is delayed and frequency modulated by a time-varying delay $\tau(t)$ and the function values of the signal are modified by the nonlinearity $f$, which is represented by the right hand side of Eq.~\eqref{eq:sys}.
According to the left hand side, the resulting signal $f(z\bm{(}R(t)\bm{)})$ is then filtered by a low-pass filter with a cutoff frequency $\Theta$ before the next roundtrip inside the feedback loop begins.
Mathematically, this process can be described by the so-called \emph{method of steps}, which is an iterative procedure for solving DDEs with constant \cite{bellman_computational_1961} and time-varying delays \cite{bellman_computational_1965}.
For that the solution is divided into suitable solution segments $z_k(t)$, with $t \in (t_{k-1},t_k] = \mathcal{I}_k$, which represent the memory of the system at time $t=t_k$.
If a signal ended a roundtip inside the feedback loop at time $t_k$, it began the roundtrip at time $t_{k-1}=R(t_k)=t_k-\tau(t_k)$.
Therefore the boundaries $t_k$ of the so-called \emph{state intervals} $\mathcal{I}_k$ of the solution segments are connected by the so-called \emph{access map} given by
\begin{equation}
	\label{eq:access_map}
	t'=R(t) = t-\tau(t),
\end{equation}
whose dynamics also plays a crucial role in mathematical properties of systems with time-varying delays \cite{mallet-paret_analyticity_2014,he_construction_2016}.
A solution segment $z_{k+1}(t)$ can be computed from the preceding segment $z_{k}(t)$ using the solution operator defined by 
\begin{equation}
	\label{eq:soluop}
	z_{k+1}(t) = z_{k}(t_{k}) e^{-\Theta(t-t_{k})} + \int\limits_{t_{k}}^{t} \! dt' \, \Theta e^{-\Theta(t-t')} f(z_k\bm{(}R(t')\bm{)}),
\end{equation}
which can be derived from Eq.~\eqref{eq:sys} by substituting the instantaneous terms $z(t)$ and $\dot{z}(t)$ with $z_{k+1}(t)$ and $\dot{z}_{k+1}(t)$, respectively, as well as the delayed term $z\bm{(}R(t)\bm{)}$ with $z_k\bm{(}R(t)\bm{)}$, and solving the resulting ordinary differential equation for $z_{k+1}(t)$.
In the limit $\Theta\to\infty$, which means that the cutoff frequency of the low-pass filter is sent to infinity, the integral kernel in Eq.~\eqref{eq:soluop} converges to a delta distribution \cite{ikeda_high-dimensional_1987} and we obtain the singular limit map
\begin{equation}
	\label{eq:limit_map}
	z_{k+1}(t) = f(z_k\bm{(}R(t)\bm{)}),
\end{equation}
which can be used to approximate the dynamics of Eq.~\eqref{eq:sys} for large $\Theta$ given that the derivative $\dot{z}(t)$ is much smaller than $\Theta$, i.e., that the characteristic timescale of the temporal structures of $z(t)$ must be coarser than the width $\Theta^{-1}$ of the kernel in Eq.~\eqref{eq:soluop} for a good approximation.
Equation~\eqref{eq:limit_map} can be interpreted as iteration of the graph $(t,z_k(t))$ under the two-dimensional map
\begin{subequations}
	\label{eq:2d_map}
	\begin{align}
		x_k &= R^{-1}(x_{k-1}) \label{eq:2d_map_r}\\
		y_k &= f(y_{k-1}) \label{eq:2d_map_f},
	\end{align}
\end{subequations}
which consists of two independent one-dimensional iterated maps, where Eq.~\eqref{eq:2d_map_r} describes the frequency modulation of the signal $z(t)$ due to the delay variation and Eq.~\eqref{eq:2d_map_f} describes the influence of the nonlinearity, which modifies the function values of $z(t)$.
Given that the map defined by Eq.~\eqref{eq:2d_map_f} shows chaos and $\Theta$ is sufficiently large, it was demonstrated in \cite{muller_laminar_2018,muller-bender_resonant_2019} that Eq.~\eqref{eq:sys} with a periodically time-varying delay shows two types of chaotic dynamics, which depend on the dynamics of the access map, Eq.~\eqref{eq:access_map}.
For periodically time-varying delay this map is a lift of the monotonically increasing circle map (cf. \cite{katok_introduction_1997})
\begin{equation}
	\label{eq:circle_map}
	\theta' = \nu_1\, R(\theta/\nu_1) \mod 1.
\end{equation}
This map shows two types of dynamics, which naturally induce two classes of time-varying delays, which lead to two types of chaos that are observed in Eq.~\eqref{eq:sys}.
For so-called \emph{conservative delay}, the circle map and its inverse defined by Eq.~\eqref{eq:2d_map_r} show quasiperiodic dynamics with a Lyapunov exponent $\lambda[R]=0$.
In this case, Eq.~\eqref{eq:2d_map_r} only leads to a quasiperiodic frequency modulation.
Stretching and folding of the chaotic map given by Eq.~\eqref{eq:2d_map_f} causes strong fluctuations, where the characteristic frequency is bounded due to the low-pass filter of the feedback loop.
The resulting dynamics is called \emph{turbulent chaos} adapted from the term ``optical turbulence'' which was introduced in \cite{ikeda_optical_1980}, where a optical feedback loop with the structure of Eq.~\eqref{eq:sys} with a constant delay was investigated.
An exemplary time series is shown in Fig.~\ref{fig:turbchaos}(a).
For so-called \emph{dissipative delays}, where the circle map associated to the access map, Eq.~\eqref{eq:access_map}, shows stable periodic dynamics with a negative Lyapunov exponent $\lambda[R]<0$.
This means we have a resonance between the average roundtrip frequency of the signal inside the feedback loop and the frequency of the time-varying delay.
This \emph{resonant Doppler effect} leads to periodically alternating phases with low and high frequencies in the solution $z(t)$.
If the condition
\begin{equation}
	\label{eq:lamchaos_crit}
	\lambda[f] + \lambda[R] < 0
\end{equation}
is fulfilled, the low-frequency regions degenerate to nearly constant laminar phases, which are periodically interrupted by irregular bursts at stable periodic points of $R^{-1} \mod 1$ corresponding to drifting stable orbits in the lift, Eq.~\eqref{eq:2d_map_r}.
This type of dynamics is called \emph{laminar chaos}.
The intensity levels of the laminar phases are determined by the dynamics of the one-dimensional map given by Eq.~\eqref{eq:2d_map_f}. 
The criterion for laminar chaos, Eq.~\eqref{eq:lamchaos_crit}, can be derived by a stability analysis of square wave solutions of the limit map, Eq.~\eqref{eq:limit_map}, cf. \cite{muller_laminar_2018}.

\section{Laminar chaos in systems with quasiperiodic delay}
\label{sec:lamchaos_quasi}

\subsection{Numerical experiments}
\label{sec:lamchaos_quasi_num}

\begin{figure}
	\includegraphics[width=0.48\textwidth]{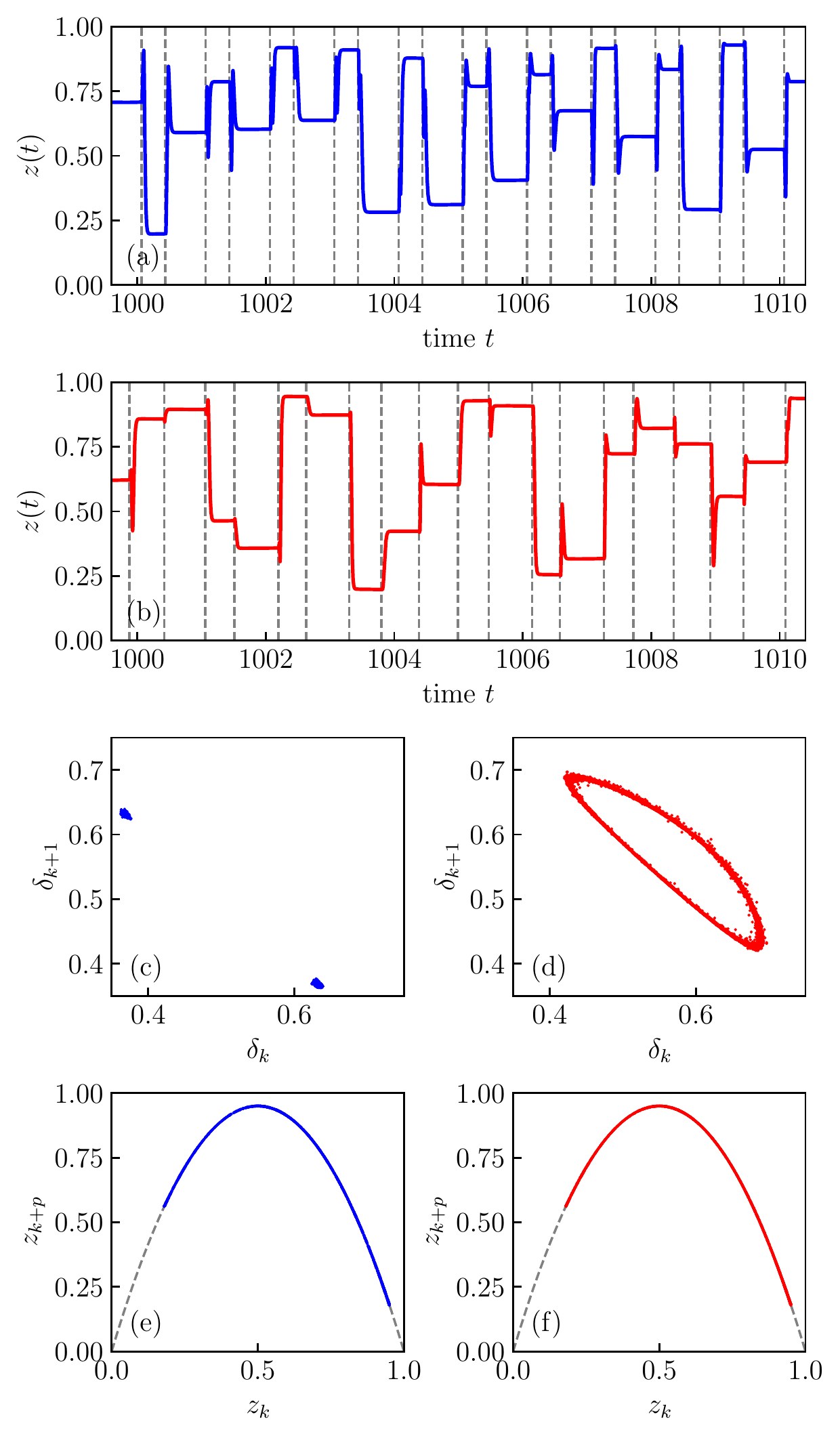}
	\caption{Features of laminar chaotic time series in systems with (a) periodic delay and (b) quasiperiodic delay.
		In (c) and (d) the duration $\delta_{k+1}$ of the $(k+1)$th laminar phase is plotted over the duration $\delta_k$ of the $k$th laminar phase for a periodic and quasiperiodic delay, respectively.
		For periodic delays the points $(\delta_k,\delta_{k+1})$ accumulate at discrete points, i.e., the durations $\delta_k$ vary periodically, whereas for quasiperiodic delay the points $(\delta_k,\delta_{k+1})$ densely fill a circle, which indicates quasiperiodic dynamics.
		The dynamics of the levels $z_k$ of the laminar phases is governed by the map $z_{n+p}=f(z_n)$ as shown in (e) and (f) with $p=3$ and $p=2$ for periodic and quasiperiodic delay, respectively, where the points $(z_k,z_{k+p})$ (dots) resemble the nonlinearity $f$ (dashed line).
		The time series were obtained from Eq.~\eqref{eq:sys} with $\Theta=100$ and $f(z)=3.8\,z(1-z)$, where the delays were chosen as in Fig.~\ref{fig:delay} with $A=0.9$ as well as $\tau_0=1.5$ for periodic and $\tau_0=1.135$ for quasiperiodic delay.
		The durations $\delta_k$ and the levels $z_k$ were estimated directly from the time series.
		We first detected the bursts times.
		Then the durations $\delta_k$ are simply the waiting times between two subsequent bursts and the levels $z_k$ are the values of the time series in the middle between two bursts.
	}
	\label{fig:lamchaos}
\end{figure}

\begin{figure}
	\includegraphics[width=0.48\textwidth]{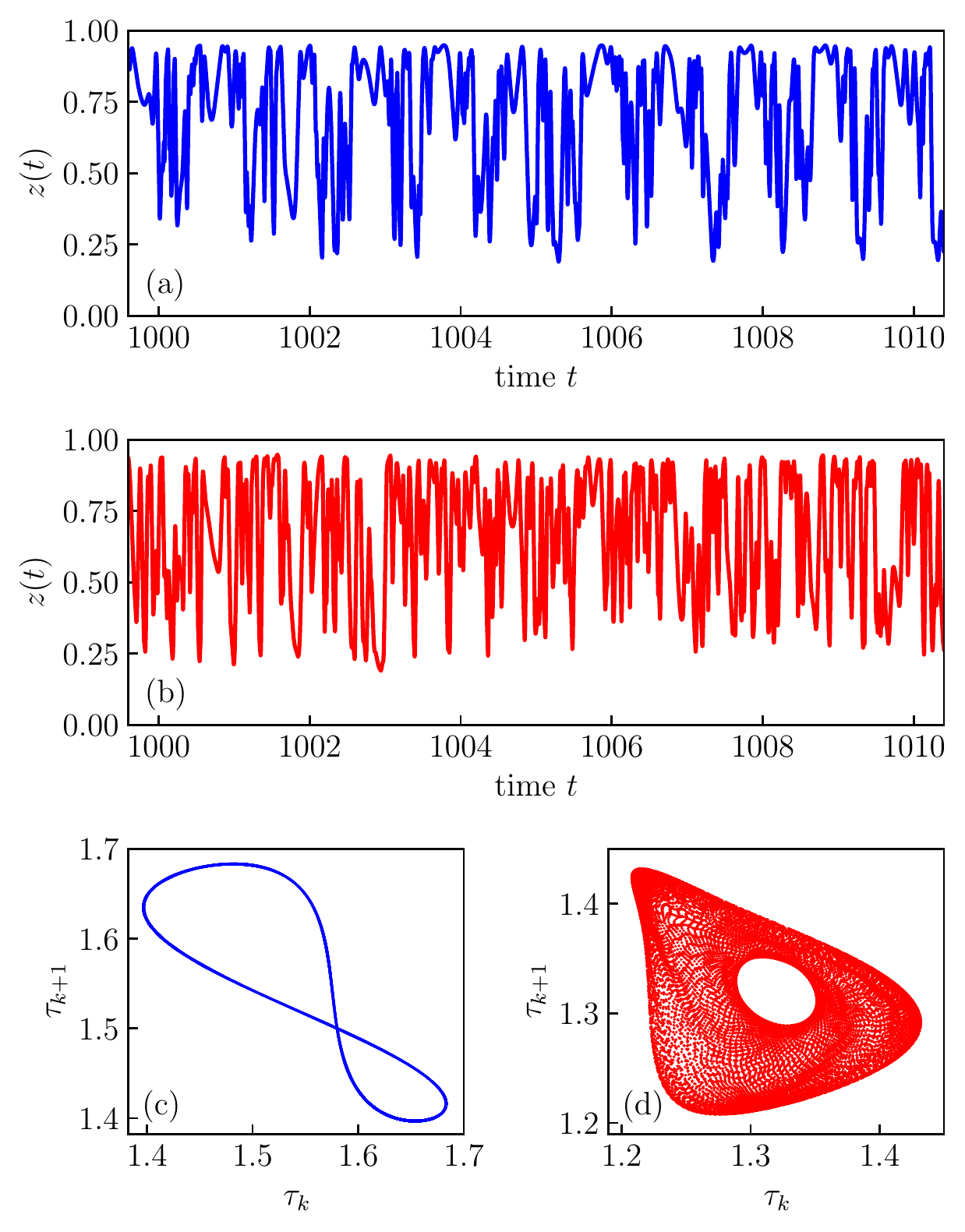}
	\caption{Features of turbulent chaotic time series in systems with (a) periodic and (b) quasiperiodic delay.
		They are characterized by chaotic fluctuations with a large frequency, which is bounded by the cutoff-frequency $\Theta$ of the low-pass filter in Eq.~\eqref{eq:sys}. 
		In (c) and (d) the length $\tau_{k+1}$ of the $(k+1)$th state interval $\mathcal{I}_{k+1} = (t_k,t_{k+1}=R^{-1}(t_k)]$ according to the method of steps, Eq.~\eqref{eq:soluop}, is plotted over the duration $\tau_k$ of the $k$th one for a periodic and a quasiperiodic delay, respectively, where we set $t_0=0$.
		For periodic delay, the dynamics of the state intervals is conjugate to a rotation on a circle, whereas for quasiperiodic delay it is conjugate to a translation on the torus.
		The time series were computed from Eq.~\eqref{eq:sys} with $\Theta=100$ and $f(z)=3.8\,z(1-z)$. We have chosen the delays as in Fig.~\ref{fig:delay} with $A=0.9$ as well as $\tau_0=1.54$ for periodic and $\tau_0=1.32$ for quasiperiodic delay, respectively.
	}
	\label{fig:turbchaos}
\end{figure}

A good starting point for the generalization of the theory of laminar chaos to systems with quasiperiodic delay is Eq.~\eqref{eq:lamchaos_crit}.
We will argue below that the Lyapunov exponent $\lambda[R]$ of the access map is well defined also for quasiperiodic delay and we naively assume that laminar chaos can be observed for large enough $\Theta$ if Eq.~\eqref{eq:lamchaos_crit} is fulfilled.
As an exemplary system we choose Eq.~\eqref{eq:sys} with the nonlinearity $f(z)=3.8\,z(1-z)$, and the quasiperiodic delay shown in Fig.~\ref{fig:delay}(b) with $A=0.9$ and $\tau_0=1.135$.
For these parameters, Eq.~\eqref{eq:lamchaos_crit} is fulfilled and the resulting time series, which is shown in Fig.~\ref{fig:lamchaos}(b), clearly shows features of laminar chaotic behavior.
There are nearly constant laminar phases, whose intensity levels are determined by the map given by Eq.~\eqref{eq:2d_map_f} as illustrated in Fig.~\ref{fig:lamchaos}(f), where the intensity level $z_{k+p}$ of the $(k+p)$th laminar phase is plotted over the level of the $k$th laminar phase for a large number of intensity levels, where $p>0$ was set to the smallest natural number such that the points $(z_k,z_{k+p})$ resemble a line.
Thus the time series passes the test for laminar chaos introduced in \cite{muller-bender_laminar_2020}.
The same behavior is observed for periodic delay as shown in Fig.~\ref{fig:lamchaos}(e).
While for a periodic delay the durations $\delta_k$ of the laminar phases vary periodically, as illustrated in Fig.~\ref{fig:lamchaos}(c), where the estimated durations alternate between two values, we observe a different behavior for a quasiperiodic delay.
The points $(\delta_k,\delta_{k+1})$ fill a closed curve, which indicates a quasiperiodic variation of the durations of the laminar phases.
Also turbulent chaos can be observed in systems with quasiperiodic delay.
An exemplary time series is shown in Fig.~\ref{fig:turbchaos}(b), which was generated from the same system, where only the mean delay was changed, $\tau_0=1.32$, leading to $\lambda[R] \approx 0$, which indicates the presence of a conservative delay.
Analogously to the durations of the laminar phases $\delta_k$, in Fig.~\ref{fig:turbchaos}(c) and (d), for periodic ($N=1$) and quasiperiodic delay ($N=2$), respectively, the dynamics of the lengths $\tau_k=\tau(t_k)$ of the state intervals $\mathcal{I}_k = (t_{k-1}, t_k] = (R(t_k),t_k] = (t_k-\tau(t_k),t_k]$ is illustrated by plotting $\tau_{k+1}$ over $\tau_k$ for turbulent chaos.
While for turbulent chaos it seems that the points $(\tau_k,\tau_{k+1})$ fill an $N$-dimensional torus, for laminar chaos the dynamics of the points $(\delta_k,\delta_{k+1})$ seem to lie on a $N-1$-dimensional torus, which is a circle for quasiperiodic delay with $N=2$ and a periodic orbit for periodic delay.
These results indicate that the concept of dissipative and conservative delays, which are associated to laminar and turbulent chaotic behavior, respectively, has a meaningful generalization to systems with quasiperiodic delay.
As we will see in the following, a quasiperiodic dissipative delay is characterized by a generalized resonance between the roundtrip inside the feedback loop and the quasiperiodic delay modulation.
If there is no such resonance, the delay is a conservative delay.

\subsection{Theory}
\label{sec:lamchaos_quasi_theory}

\begin{figure}
	\includegraphics[width=0.48\textwidth]{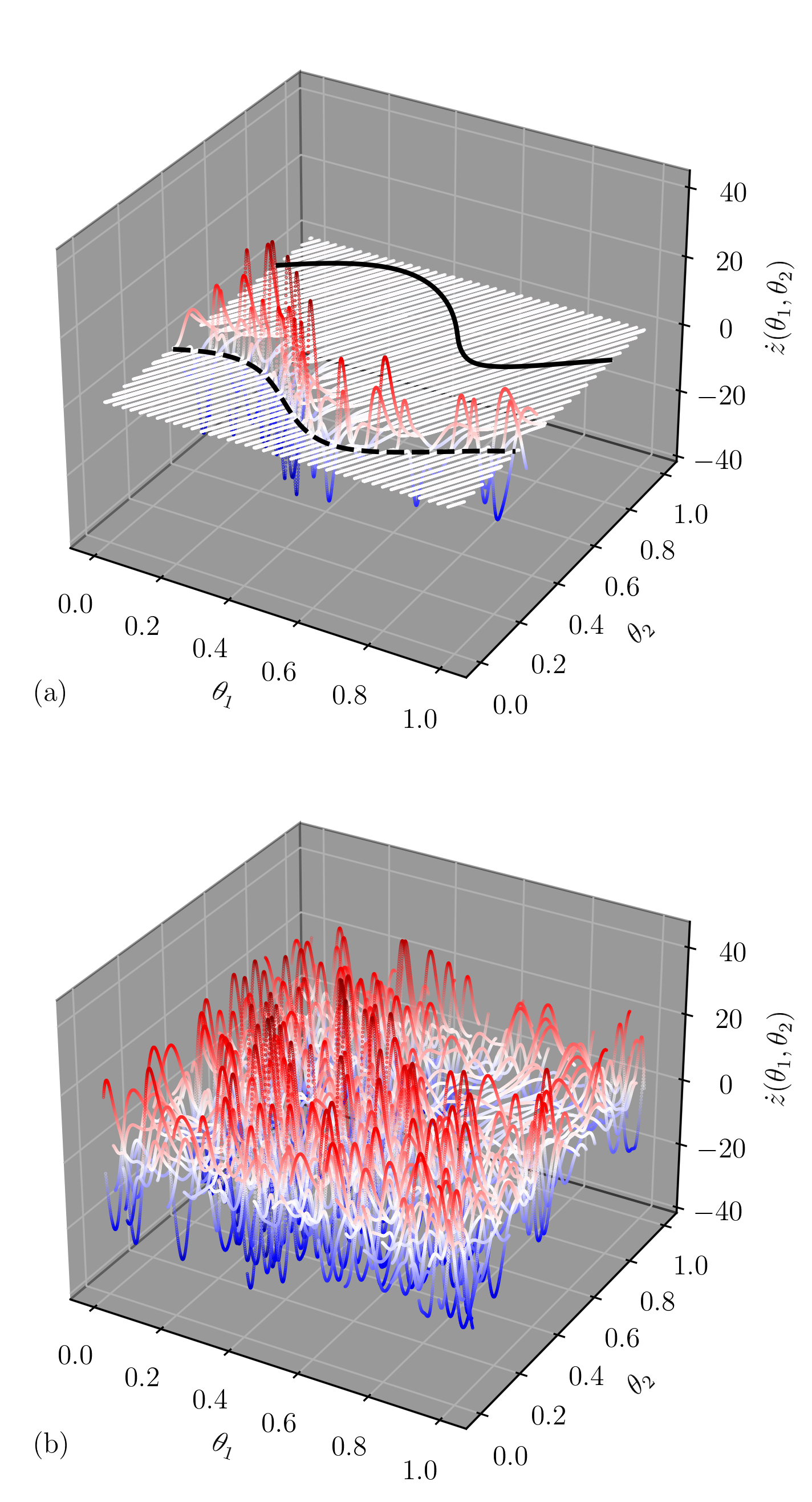}
	\caption{Properties of the dynamics governed by Eq.~\eqref{eq:sys} are revealed by the projection $\dot{z}(\theta_1,\theta_2)$ of the time series derivative $\dot{z}(t)$ to the torus map that is associated with the quasiperiodic delay for (a) laminar chaos and (b) turbulent chaos.
		The relation between the torus coordinates $\theta_n$ and time $t$ is given by Eq.~\eqref{eq:torus_embedding}.
		For laminar chaos the bursts between laminar phases accumulate at the unstable manifold (dashed line) of the torus map, Eqs.~(\ref{eq:reduced_access_map},\ref{eq:reduced_access_map_component}), whereas the stable manifold lies inside the laminar phases.
		For turbulent chaos, the quasiperiodic frequency modulation of the strongly fluctuating signal becomes roughly visible.
		For better visualization, the derivative was also color coded, where red, white, and blue, represent positive, nearly zero, and negative values, respectively.
		The parameters were chosen as in Fig.~\ref{fig:lamchaos}(b) and Fig.~\ref{fig:turbchaos}(b).
	}
	\label{fig:trajproj}
\end{figure}

\begin{figure}
	\includegraphics[width=0.47\textwidth]{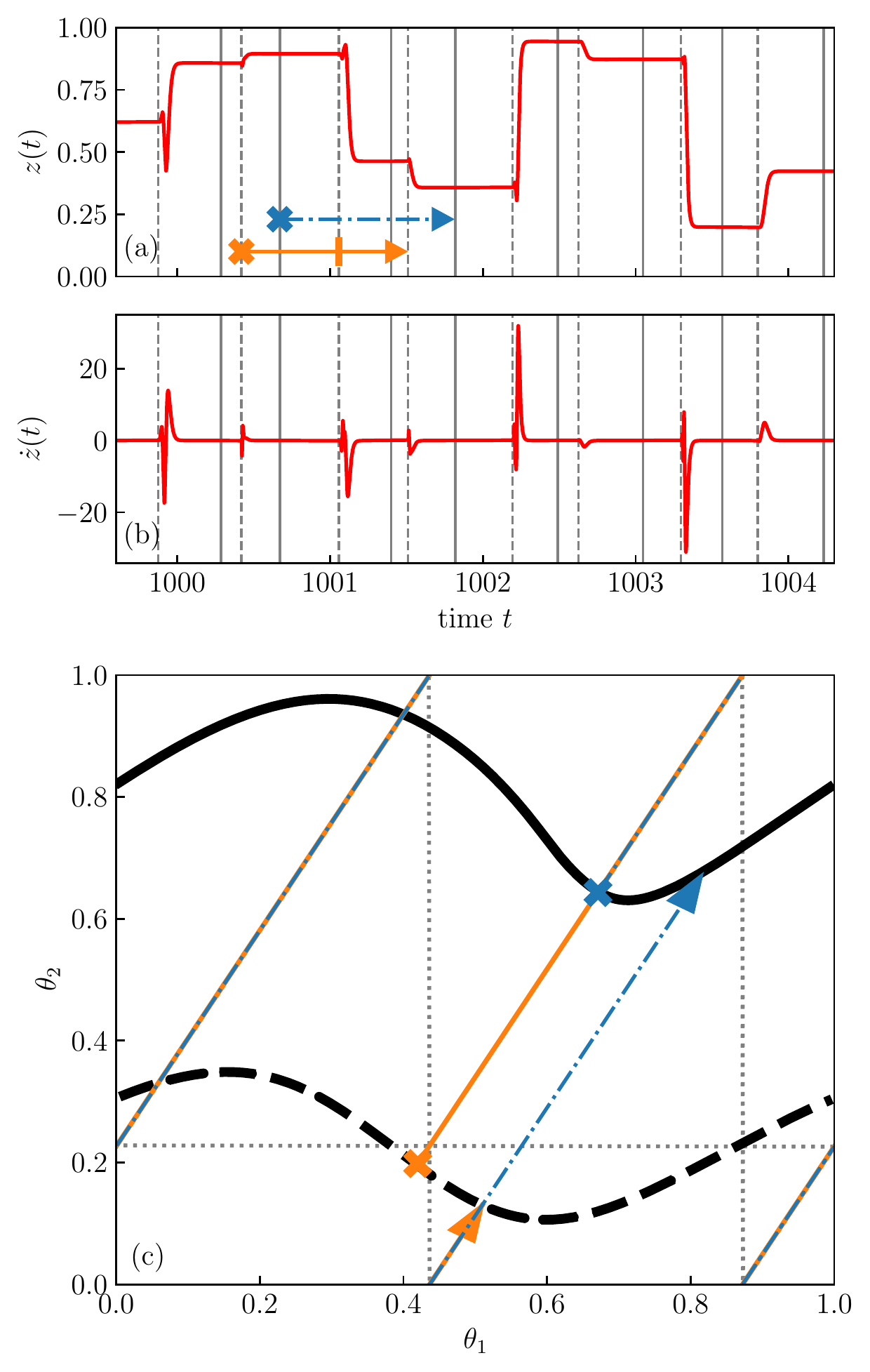}
	\caption{Mechanism behind the quasiperiodic variation of the durations and the chaotic variation of the levels of the laminar phases.
		In (a) a section of the laminar chaotic time series $z(t)$ from Fig.~\ref{fig:lamchaos}(b) is shown together with two characteristic time intervals (horizontal lines), which are of one delay length, i.e., the left boundary $t_{k-1}=R(t_k)$ (cross) is connected to the right boundary $t_k$ (arrow head) by the access map, Eq.~\eqref{eq:access_map}.
		The derivative $\dot{z}(t)$ is shown in (b).
		In (c) the intervals from (a) are projected to the toroidal phase space of the torus map, Eq.~\eqref{eq:reduced_access_map_component_example}, see Eq.~\eqref{eq:torus_embedding} and text below.
		Whenever the projection of the time-interval intersects the unstable manifold of the torus map (thick dashed line), a burst between two laminar phases is observed (dashed vertical lines in (a)).
		The intersections with the stable manifold (thick line) and the associated time instants (solid vertical lines in (a)) determine the long time dynamics in the limit $\Theta\to\infty$ given by Eq.~\eqref{eq:2d_map} (see text).
		Especially, it follows that the one-dimensional map given by Eq.~\eqref{eq:2d_map_f} governs the dynamics of the levels of the laminar phases, where the $k$th laminar phase is mapped to the $(k+p)$th laminar phase, where $p$ is the numerator of the average roundtrip time $T$ given by Eq.~\eqref{eq:roundtrip_time}.
		Here $T$ is determined by $q_1=0$, $q_2=1$, and $p=2$.
		In this case the $k$th laminar phase is mapped to the $(k+2)$th laminar phase as indicated by the boundaries of the dash-dotted interval in (a).
		The stable and unstable manifolds in (c) were computed by projecting a settled orbit of the access map or its inverse, respectively, to the toroidal phase space using Eq.~\eqref{eq:torus_embedding}.
	}
	\label{fig:access_map}
\end{figure}

\begin{figure}
	\includegraphics[width=0.48\textwidth]{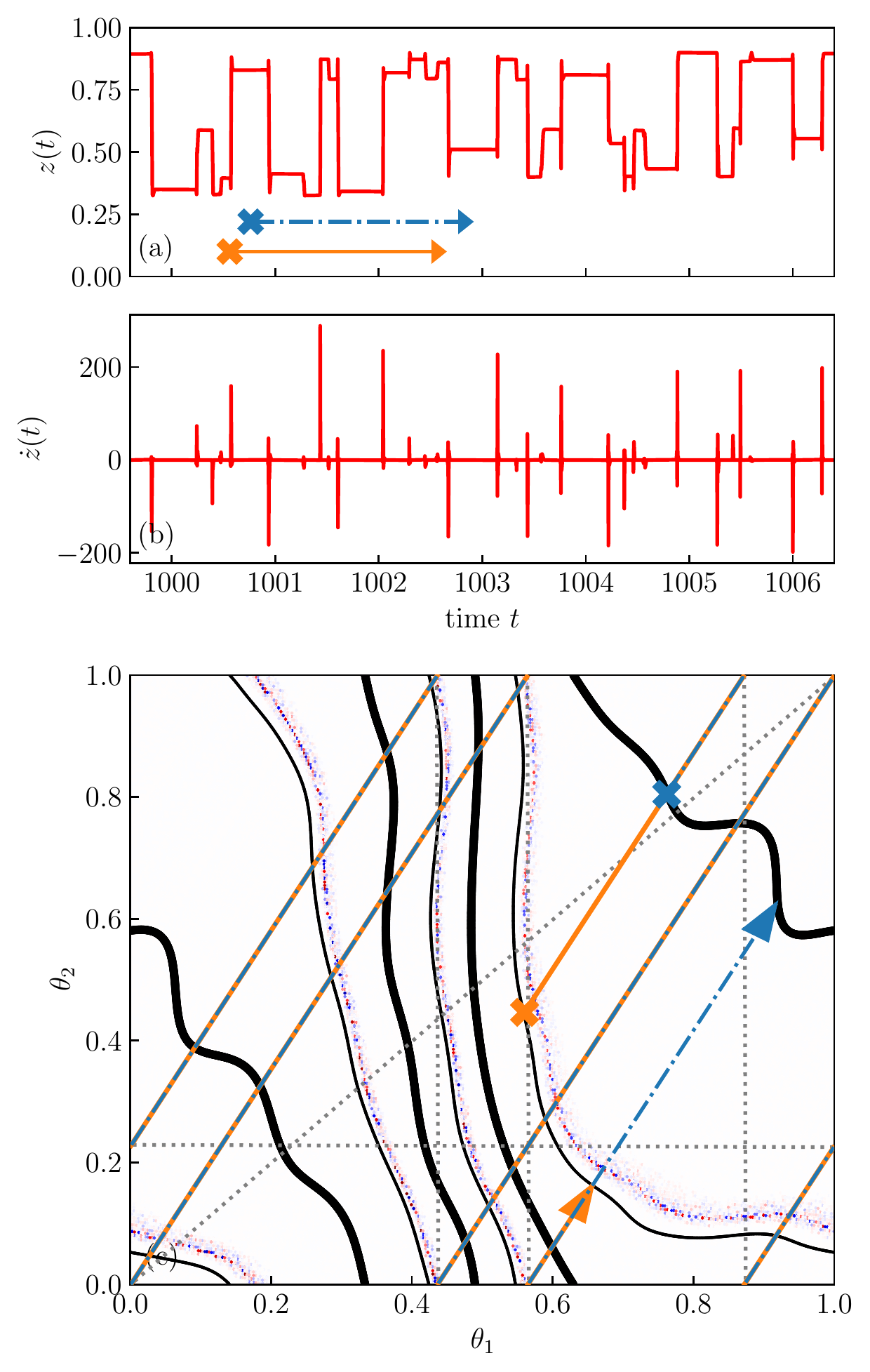}
	\caption{Laminar chaos with a more complex structure of the laminar phases.
		Here, the same analysis as in Fig.~\ref{fig:access_map} is done for Eq.~\eqref{eq:sys} with $\Theta=1000$, the nonlinearity $f(z)=3.6\,z(1-z)$, and the quasiperiodic delay Eq.~\eqref{eq:delay_def}, where we set $N=2$, $\nu_1=1$, $\nu_2=\sqrt{\pi}$, $A=1.0$, and $\tau_0=2.1183$.
		The stable and unstable manifold of the torus map Eq.~\eqref{eq:reduced_access_map_component_example} are represented by the thick and thin black lines, respectively.
		The average roundtrip time $T$, Eq.~\eqref{eq:roundtrip_time} is determined by $q_1=3$, $q_2=1$, and $p=10$.
		In (c), we additionally show the projection $\dot{z}(\theta_1,\theta_2)$ of the derivative $\dot{z}(t)$ as done in Fig.~\eqref{fig:trajproj}, where the derivative is color coded such that red, white, and blue, represent positive, nearly zero, and negative values, respectively.
		As expected from the theory, the bursts (red and blue dots) that correspond to the transitions between the laminar phases accumulate at the unstable manifold of the torus map, whereas the laminar phases (white) lie in between.
	}
	\label{fig:access_map_app}
\end{figure}

To understand the numerical results in detail we consider the dynamics of the 2d-map, Eq.~\eqref{eq:2d_map} and its connection to the dynamics of Eq.~\eqref{eq:sys}.
Since Eq.~\eqref{eq:2d_map_f} is independent of the delay, we focus on Eq.~\eqref{eq:2d_map_r}, and its inverse, the access map, Eq.~\eqref{eq:access_map}, which also holds for quasiperiodic delay.
While for a proper analysis in the case of a periodic delay, the access map can be reduced to a circle map, Eq.~\eqref{eq:circle_map}, for quasiperiodic delay, the dynamics of the access map can be mapped to the dynamics of an $N$-dimensional torus map,
\begin{equation}
	\label{eq:reduced_access_map}
	\vec{\theta}' = \vec{r}(\vec{\theta}),
\end{equation}
which is defined by the component wise representation
\begin{equation}
	\label{eq:reduced_access_map_component}
	\theta_{n}' = r_n(\vec{\theta}) = [\theta_{n} - \tau_0\nu_n - A\nu_n\,g(\theta_{1},\theta_{2},\dots,\theta_{N})] \mod 1.
\end{equation}
For our numerical examples of quasiperiodic delays with $N=2$ frequencies ($\nu_1=1$, $\nu_2=\sqrt{\pi}$) considered in this and the previous section, where $\tau(t)$ is defined by Eq.~\eqref{eq:delay_def}, we obtain
\begin{subequations}
	\label{eq:reduced_access_map_component_example}
	\begin{align}
		\theta_{1}' &= \left\{ \theta_{1} - \tau_0\nu_1 - \frac{A\nu_1}{2} \left[ \frac{\cos(2\pi \theta_{1})}{2\pi\nu_{1}} + \frac{\cos(2\pi \theta_{2})}{2\pi\nu_{2}} \right] \right\}\, \text{mod } 1 \\
		\theta_{2}' &= \left\{ \theta_{2} - \tau_0\nu_2 - \frac{A\nu_2}{2} \left[ \frac{\cos(2\pi \theta_{1})}{2\pi\nu_{1}} + \frac{\cos(2\pi \theta_{2})}{2\pi\nu_{2}} \right] \right\}\, \text{mod } 1.
	\end{align}
\end{subequations}
With the relation 
\begin{equation}
	\label{eq:torus_embedding}
	\theta_{n}=\nu_n t \mod 1,
\end{equation}
each orbit of the access map, Eq.~\eqref{eq:access_map}, corresponds to an orbit of the torus map, Eqs.~(\ref{eq:reduced_access_map},\ref{eq:reduced_access_map_component}) and, given that the frequencies $\nu_n$ are incommensurate, also vice versa.
The mapping defined by Eq.~\eqref{eq:torus_embedding} consists of two steps:
First, the value $t$ from the domain $\mathbb{R}$ of the access map, Eq.~\eqref{eq:access_map}, is embedded into $\mathbb{R}^N$ by the linear mapping $t \to (\nu_1 t, \nu_2 t, \dots, \nu_N t)^T$.
The resulting $N$-dimensional vector is then projected to the $N$-dimensional torus $\mathbb{T}^N$, which is the domain of the torus map, Eqs.~(\ref{eq:reduced_access_map},\ref{eq:reduced_access_map_component}), by taking each component modulo one.
By applying Eq.~\eqref{eq:torus_embedding} to all times $t$ of a state interval $\mathcal{I}_k$ or other subsets of the time domain of the delay system, Eq.~\eqref{eq:sys}, these subsets can be projected to the domain of the torus map as done in Fig.~\ref{fig:trajproj}, Fig.~\ref{fig:access_map}, and Fig.~\ref{fig:access_map_app}, which enables the analysis of the connection between dynamical properties of the torus map and the dynamics of the delay system.
In Fig.~\ref{fig:trajproj}, the derivative $\dot{z}(t)$ of an exemplary time series $z(t)$ of Eq.~\eqref{eq:sys} with an $N=2$ frequency quasiperiodic delay is shown over the torus $\mathbb{T}^2$, which is the domain of the related torus map, Eq.~\eqref{eq:reduced_access_map_component_example}.
For that we plotted $\dot{z}(\theta_1,\theta_2)=\dot{z}(t)$ over the $\theta_1$-$\theta_2$ plane, where the $\theta_n$ are computed from time $t$ using Eq.~\eqref{eq:torus_embedding}.
For laminar chaos, Fig.~\ref{fig:trajproj}(a), the nearly constant laminar phases correspond to parallel lines in the $\dot{z}=0$ plane, and the transitions between them correspond to large amplitude bursts.
These bursts accumulate at an $N-1=1$ dimensional manifold, which is the unstable manifold of the torus map as shown below.
In contrast, for turbulent chaos, which is shown in Fig.~\ref{fig:trajproj}(b), strong fluctuations cover the whole $\theta_1$-$\theta_2$ plane, where frequency and amplitude of the derivative appear to vary with the position on the plane, which indicates the frequency modulation of the turbulent chaotic signal due to the non-resonant Doppler effect, which is also observed in systems with periodic delay \cite{muller-bender_resonant_2019}.
In the following, these results are explained in detail, where we discuss the dynamics of the torus map, Eqs.~(\ref{eq:reduced_access_map},\ref{eq:reduced_access_map_component}), and reveal its influence to the dynamics of the delay system, Eq.~\eqref{eq:sys}.

Torus maps that are equivalent to a one-dimensional quasiperiodic map belong to the class of \emph{foliation preserving torus maps} since they preserve the foliation defined by the projection of the domain $\mathbb{R}$ of the one-dimensional map to the domain $\mathbb{T}^N$ of the torus map via Eq.~\eqref{eq:torus_embedding}.
Such maps were extensively analyzed in \cite{petrov_torus_2003}, where the one-dimensional map models the reflections of light in an optical resonator with quasiperiodically moving walls.
Further mathematical results can be found in \cite{he_resonances_2022}.
An important quantity for characterizing the dynamics of the feedback loop, Eq.~\eqref{eq:sys}, and the associated maps is the average roundtrip time $T=v[R^{-1}]=-v[R]$, where $v[R]$ is the drift velocity defined by
\begin{equation}
	\label{eq:drift_velocity}
	v[R] = \lim_{k\to\infty} \frac{R^k(t_0) - t_0}{k}.
\end{equation}
Another characteristic quantity is the so-called rotation vector $\vec{\rho}[\vec{r}]$ (cf. \cite{petrov_torus_2003}).
Its components $\rho_n$ are the average number of rotations in each direction of the torus per iteration of the torus map, Eqs.~(\ref{eq:reduced_access_map},\ref{eq:reduced_access_map_component}) and are given by
\begin{equation}
	\label{eq:rotation_vector}
	\rho_n = \lim_{K\to\infty} -\frac{\nu_n}{K} \sum_{k=0}^{K-1} [\tau_0 + A\,g(\theta_{k,1},\theta_{k,2},\dots,\theta_{k,N})] = \nu_n v[R],
\end{equation}
which is the average increment of the $n$th component of the torus map over a reference orbit $\{\vec{\theta}_k\}_{k\in\mathbb{N}}$.
The Lyapunov exponent $\lambda[R]$ of the access map given by \cite{ott_chaos_2002}
\begin{equation}
	\label{eq:lyapunov_exponent}
	\lambda[R] = \lim_{K\to\infty} \frac{1}{K} \sum_{k=0}^{K-1} \ln|R'(R^k(t_0))|
\end{equation}
is equal to the Lyapunov exponent of the torus map that is associated to perturbations in the direction of the frequency vector $(\nu_1,\nu_2,\dots,\nu_N)^{\mathrm{T}}$, while all other exponents are equal to zero \cite{petrov_torus_2003}. 
For periodic delay, $N=1$, where the torus map degenerates to the circle map given by Eq.~\eqref{eq:circle_map}, the rotation vector has only one component $\rho_1=\nu_1 v[R]$, which is the so-called rotation number (cf. \cite{katok_introduction_1997}).
While an irrational rotation number implies quasiperiodic dynamics with $\lambda[R]=0$ \cite{de_faria_real_2016}, mode-locking dynamics with $\lambda[R]<0$ is characterized by a rational rotation number, where the denominator is equal to the period of the stable periodic orbit.
For quasiperiodic delay, the mode-locking dynamics known from the circle map is replaced by generalized mode-locking, where we have $\lambda[R]<0$ as in the case of a periodic delay but the average roundtrip time $T$ now has the structure \cite{petrov_torus_2003}
\begin{equation}
	\label{eq:roundtrip_time}
	T = \frac{p}{\sum_{n=1}^{N} q_n\,\nu_n},
\end{equation}
where $p$ and the $q_n$ are integers, i.e., the average roundrip frequency $T^{-1}$ inside the feedback loop is a sum of rational multiples of the delay modulation frequencies.
In other words, the components of the rotation vector $\vec{\rho}$ fulfill the resonance condition
\begin{equation}
	p + \sum_{n=1}^{N} q_n\,\rho_n = 0.
\end{equation}
It follows, for dissipative quasiperiodic delay, there is a generalized resonance between the roundtrip inside the feedback loop and the delay modulation similar to the resonance observed for periodic delay in \cite{muller-bender_resonant_2019}.
Since the stable and the unstable manifolds are homotopic to the set $\{ \vec{\theta} \cdot \vec{q} = 0 \;|\; \vec{\theta} \in \mathbb{T}^N \}$, where we have $\vec{q}=(q_1,q_2,\dots,q_N)^T$ \cite{petrov_torus_2003}, for $N=2$, the integers $q_n$ and $p$ can be obtained by a graphical analysis of the structure of the stable and unstable manifolds of the torus map, which are represented by the black solid and dashed lines of Fig.~\ref{fig:access_map} or the black thick and thin lines of Fig.~\ref{fig:access_map_app}, respectively.
The moduli of the $q_n$ are given by the number of unidirectional intersections of such a manifold with the $\theta_n$ axis, and $p$ is the number of intersections of the projection of the interval $(R(t),t]$ to the toroidal phase space by Eq.~\eqref{eq:torus_embedding} with the stable or unstable manifold.
Such intervals $(R(t),t]$ and their projection are shown in Fig.~\ref{fig:access_map} and Fig.~\ref{fig:access_map_app}.
The signs of the $q_n$ must be chosen such that Eq.~\eqref{eq:roundtrip_time} is fulfilled.

Under the given assumptions of incommensurate frequencies $\nu_1,\nu_2,\dots,\nu_N$ and $0\leq A\leq 1$ such that the access map is monotonically increasing, the access map, Eq.~\eqref{eq:access_map}, and the associated torus map, Eqs.~(\ref{eq:reduced_access_map},\ref{eq:reduced_access_map_component}), basically show two types of dynamics \cite{petrov_torus_2003}, which define two types of quasiperiodically time-varying delays, which lead to different types of chaos in Eq.~\eqref{eq:sys} as already known for periodic delay and as indicated by the above results.
For dissipative delay the Lyapunov exponent $\lambda[R]$ of the access map is negative.
It follows that the torus map has one negative Lyapunov exponent and all other Lyapunov exponents are equal to zero.
The torus map shows generalized mode-locking, which means that almost all orbits of Eqs.~(\ref{eq:reduced_access_map},\ref{eq:reduced_access_map_component}) are attracted by a stable orbit that fills an $(N-1)$-dimensional torus.
This stable manifold is represented by the thick solid line in Fig.~\ref{fig:trajproj}(a) and Fig.~\ref{fig:access_map}(c).
Under the inverse of the torus map that is associated to Eq.~\eqref{eq:2d_map_r} almost all orbits are attracted by the unstable manifold of Eqs.~(\ref{eq:reduced_access_map},\ref{eq:reduced_access_map_component}), which is represented by the thick dashed line in Fig.~\ref{fig:trajproj}(a) and Fig.~\ref{fig:access_map}(c).
By Eq.~\eqref{eq:torus_embedding} the stable and unstable manifolds of Eqs.~(\ref{eq:reduced_access_map},\ref{eq:reduced_access_map_component}) are directly connected to the attractive and repulsive points of the access map, which determine the positions of the laminar phases and the transitions in between, respectively.
This can be made plausible by projecting the state intervals $\mathcal{I}_k = (t_{k-1},t_k]=(R(t_k),t_k]=(t_k-\tau(t_k),t_k]$ from the method of steps to the state space of the torus map as done in Fig.~\ref{fig:access_map}.
The intervals and their projections are represented by the colored lines, where the left and right interval boundaries are marked by crosses and arrow heads, respectively.
If one of the interval boundaries coincides with an attractive or repulsive point of the access map, Eq.~\eqref{eq:access_map}, which are represented by the solid and dashed vertical lines in Fig.~\ref{fig:access_map}(a), the other boundary also coincides with an attractive or repulsive point, respectively.
Whenever the projection of a state interval intersects the unstable manifold in Fig.~\ref{fig:access_map}(c), a burst-like transition between two laminar phases is observed in Fig.~\ref{fig:access_map}(a).
In the following we assume that the boundaries $t_k$ of the state intervals coincide with the repulsive points of the access map as the solidly marked interval in Fig.~\ref{fig:access_map}, which is reasonable since for almost all $t_0$, the boundaries get arbitrarily close to these points in the limit $k\to\infty$ since they evolve according to the inverse access map, Eq.~\eqref{eq:2d_map_r}.
As it was found for periodic delay \cite{muller-bender_resonant_2019}, then each state interval consists of $p$ subintervals $u_{k,i}$, $i=1,2,\dots,p$, where $p$ is the numerator of the average roundtrip time for dissipative delay given by Eq.~\eqref{eq:roundtrip_time}.
In Fig.~\ref{fig:access_map}(a) the subintervals of the solidly marked interval are separated by the vertical thick line.
The subintervals of subsequent state intervals $\mathcal{I}_k$ are connected by $u_{k-1,i}=R(u_{k,i})$, and their boundaries are the repulsive points of the access map.
If the condition for laminar chaos, Eq.~\eqref{eq:lamchaos_crit} is fulfilled, the $u_{k,i}$ are the time-domains of the laminar phases.
The $u_{k,i}$ are the basins of attractions of the attractive points of the access map, where the attractive points are the inverse images of the intersections of the projected state intervals with the stable manifold of the torus map.
The attractive points are always located inside the laminar phases and are crucial for the dynamics of the intensity levels of the laminar phases under Eq.~\eqref{eq:2d_map}.
In detail, the solution segment $z_k(t)$ is connected to the initial solution segment $z_0(t)$ with $t\in (t_{-1},t_0]$ by the $k$th iteration of the limit map, Eq.~\eqref{eq:limit_map}, which gives
\begin{equation}
	\label{eq:limit_map_k}
	z_k(t) = f^k(z_0(R^k(t))).
\end{equation}
If $t \in u_{k,i}$ is not equal to a repulsive point of the access map, the $k$th iteration $R^k(t)$ of the access map converges to an attractive point $t^*_i\in u_{0,i}$ of the initial state interval $\mathcal{I}_0$ in the limit $k\to\infty$ and we have $z_k(t) \approx f^k(z_0(t^*_i))$, which implies that the intensity levels of the laminar phases follow the dynamics of Eq.~\eqref{eq:2d_map_f} as visualized in Fig.~\ref{fig:lamchaos}(e) and (f).
The durations $\delta_k$ of the laminar phases, which are simply the lengths of the subintervals $u_{k,i}$ vary quasiperiodically since the projections of the boundaries of the $u_{k,i}$ to the phase space of the torus map, Eqs.~(\ref{eq:reduced_access_map},\ref{eq:reduced_access_map_component}), show quasiperiodic dynamics on the unstable manifold.
A similar behavior can be observed for the lengths $\tau_k=\tau(t_k)$ of the state intervals $\mathcal{I}_k = (t_{k-1}, t_k] = (R(t_k),t_k] = (t_k-\tau(t_k),t_k]$ since $\delta_k$ and $\tau_k$ are connected by $\tau_k = \sum_{k'=0}^{p-1} \delta_{k p + k'}$.
That Eq.~\eqref{eq:lamchaos_crit} is a necessary condition for laminar chaos also in the case of a quasiperiodic delay can easily be shown by analyzing Eq.~\eqref{eq:limit_map_k} in the same way as it was done for periodic delay in \cite{muller_laminar_2018}.
Laminar phases can develop if the first derivative of the right hand side of Eq.~\eqref{eq:limit_map_k} vanishes in the limit $k\to\infty$ for almost all $t\in (t_{k-1},t_k]$, which requires that Eq.~\eqref{eq:lamchaos_crit} is fulfilled.
In Fig.~\ref{fig:access_map_app}, the same analysis as in Fig.~\ref{fig:access_map} is performed for another parameter set, which is associated with a more complicated structure of the stable and unstable manifolds of the torus map, which leads to a more pronounced variation of the durations of the laminar phases.
We added the projection $\dot{z}(\theta_1,\theta_2)$ of the derivative $\dot{z}(t)$ of a related laminar chaotic time series to Fig.~\ref{fig:access_map_app}, which confirms the result that the transitions between the laminar phases correspond to the crossing points of the lines given by Eq.~\eqref{eq:torus_embedding} with the unstable manifold of the torus map.
The slight deviation is caused by the asymmetric kernel of the solution operator, Eq.~\eqref{eq:soluop}, which can be interpreted as a slight shift of the access map in negative direction as known from systems with periodic delay \cite{muller-bender_resonant_2019}.
Roughly speaking, the system ``sees'' a slightly larger delay.
Further examples for the structure of stable and unstable manifolds can be found in \cite{petrov_torus_2003}.

In strong contrast, for conservative delays, we have $\lambda[R]=0$, which implies that all Lyapunov exponents of the torus map Eqs.~(\ref{eq:reduced_access_map},\ref{eq:reduced_access_map_component}) are equal to zero.
Furthermore, there is no generalized resonance between the roundtrip inside the feedback loop and the quasiperiodic delay modulation, i.e., the average roundtrip time $T$ does not fulfill Eq.~\eqref{eq:roundtrip_time}.
In this case the torus map given by Eqs.~(\ref{eq:reduced_access_map},\ref{eq:reduced_access_map_component}) has no attractor and all orbits fill the whole $N$-dimensional torus, which implies that the access map has no attractive and repulsive points.
As demonstrated in Appendix~\ref{sec:app_trafo}, Eq.~\eqref{eq:sys} then can be transformed to a system with constant delay, where only turbulent chaos can be observed.
The dynamics of the torus map is visualized by the lengths $\tau_k$ of the state intervals $\mathcal{I}_k$ in Fig.~\ref{fig:turbchaos}(c) and (d) for a periodic conservative delay and a quasiperiodic conservative delay with $N=2$ frequencies.
While for the periodic delay the points $(\tau_k,\tau_{k+1})$ resemble a circle, for quasiperiodic delay they resemble a two-dimensional torus.
In strong contrast to laminar chaos, this dynamics can not be reconstructed easily from the time series.
While for laminar chaos one simply needs to detect the transitions between the laminar phases, the detection of the lengths of the state intervals for turbulent chaos in systems with time-varying delay is nontrivial \cite{kye_characteristics_2004}.
Therefore we computed the results in Fig.~\ref{fig:turbchaos}(c) and (d) directly from Eq.~\eqref{eq:2d_map_r}.

\section{Dimension of chaotic attractors in systems with quasiperiodic delay}
\label{sec:kydim}

\begin{figure*}
	\includegraphics[width=0.99\textwidth]{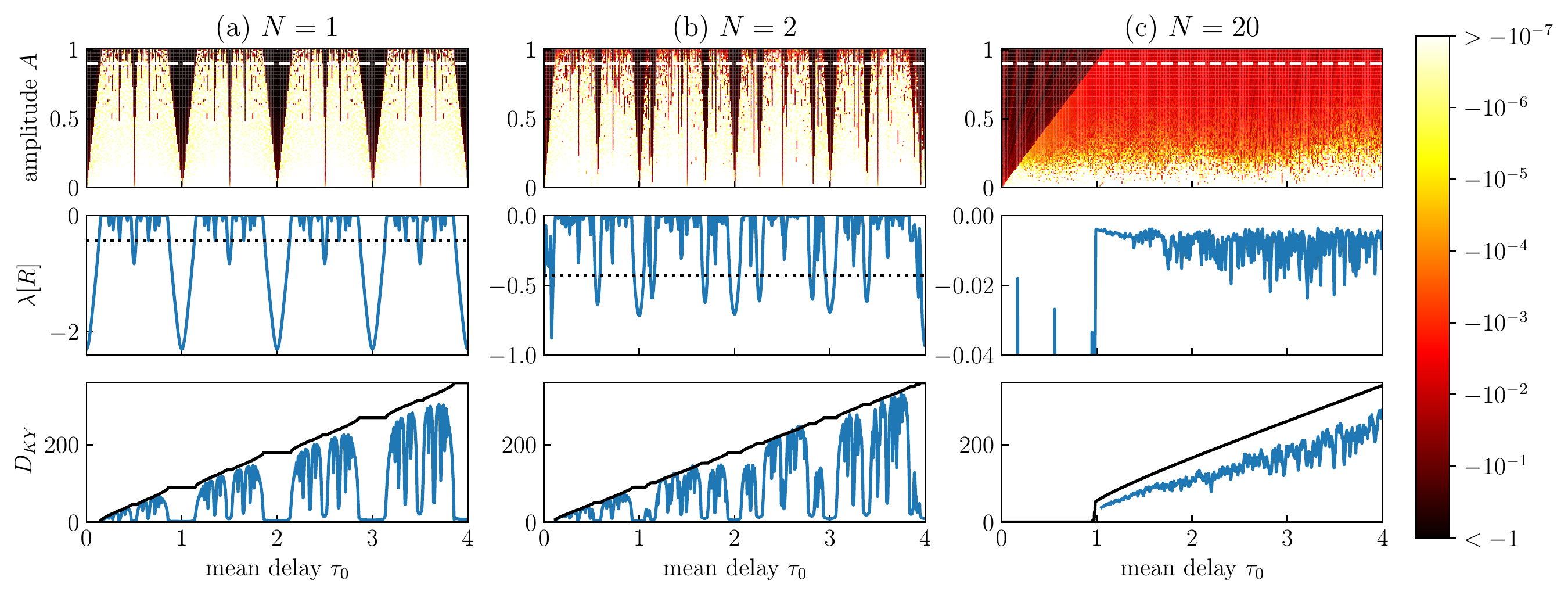}
	\caption{Influence of the time-varying delay given by Eq.~\eqref{eq:delay_def} on the dimension of the chaotic attractors of Eq.~\eqref{eq:sys} for (a) a periodic delay ($N=1$) (b) a quasiperiodic delay with $N=2$ frequencies $\nu_1=1$ and $\nu_2=\sqrt{\pi}$, and (c) a quasiperiodic delay with $N=20$ frequencies that are chosen randomly from a Gaussian distribution with zero mean and standard deviation $\sigma=2$.
		In the top panels, heat maps of the Lyapunov exponent $\lambda[R]$ of the access map, Eq.~\eqref{eq:access_map}, with respect to the mean delay $\tau_0$ and the delay amplitude $A$ are shown.
		In the center panels $\lambda[R]$ is plotted as a function $\tau_0$ for a fixed amplitude $A=0.9$ (white dashed line in the top panels).
		For $A=0.9$ the Kaplan-Yorke dimension of the chaotic attractors of Eq.~\eqref{eq:sys} with $\Theta=200$ and with the feedback nonlinearity $f(z)=3.8\,z(1-z)$ as a function of $\tau_0$ is shown in the bottom panels.
		The condition, Eq.~\eqref{eq:lamchaos_crit}, is fulfilled below the dotted line.
		Dissipative delays, $\lambda[R]<0$, lead to a giant reduction of the Kaplan-Yorke dimension compared to conservative delays including constant delays, where we have $\lambda[R]=0$.
		The periodic and quasiperiodic variation of $\lambda[R]$ with respect to $\tau_0$ leads to a periodic and quasiperiodic modulation of the Kaplan-Yorke dimension $D_{KY}$.
		For a larger number of frequencies (c) the Lyapunov exponent $\lambda[R]$ seems to be negative for a large part of the parameter space leading to an overall reduction of $D_{KY}$ compared to constant delay.
		The black solid lines represent estimates of $D_{KY}$ for a system, where the time-varying delay is replaced by a constant delay $\tau(t)=-v[R]$, where $v[R]$ is the drift velocity of the access map, while keeping all other parameters fixed (see text).
	}
	\label{fig:arnoldkydim}
\end{figure*}

As shown for periodic delays in \cite{muller_laminar_2018,muller-bender_resonant_2019} a time-varying delay can have a large influence on the dimension of chaotic attractors of time-delay systems, where the dimension depends strongly on the delay type and, therefore, it depends strongly on the delay parameters.
For conservative delays, Eq.~\eqref{eq:sys} shows turbulent chaos.
Since systems with conservative delays can be transformed to systems with constant delay one can adapt the results from \cite{farmer_chaotic_1982}, where it was demonstrated that the Kaplan-Yorke dimension $D_{KY}$ of turbulent chaos increases proportionally to the duration of the constant delay, which is equivalent to a linear growth with $\Theta$ in Eq.~\eqref{eq:sys}.
In this sense, turbulent chaos is a very high-dimensional phenomenon.
In contrast, laminar chaos is a very low-dimensional phenomenon \cite{muller_laminar_2018,muller-bender_resonant_2019}, which can be made plausible even by looking at the time series.
While turbulent chaos shows chaotic high-frequent oscillations, where the state interval must be sampled by a large number of points to describe the system memory adequately, one can expect that the memory of a system showing laminar chaos can be well described by the intensity levels of the laminar phases inside one delay interval.
This argumentation can be generalized to general dissipative delays, where we have $\lambda[R]<0$ but Eq.~\eqref{eq:lamchaos_crit} is not necessarily fulfilled.
Therefore, in \cite{muller-bender_resonant_2019} so-called generalized laminar chaos was introduced, which characterizes the dynamics of Eq.~\eqref{eq:sys} with a dissipative delay.
With the results from Sec.~\ref{sec:lamchaos_quasi}, the results for periodic delay can be applied directly to systems with quasiperiodic delay.
Generalized laminar chaos of order $m-1$ is observed, if $m>0$ is the smallest integer such that
\begin{equation}
	\label{eq:genlamchaos_crit}
	\lambda[f] + m\,\lambda[R] < 0,
\end{equation}
is fulfilled.
A stability analysis of the limit map, Eq.~\eqref{eq:limit_map}, in \cite{muller-bender_resonant_2019} reveals that the unstable directions of the limit map dynamics are given by polynomials of order $m-1$, which includes classical laminar chaos, where the unstable directions are given by zeroth order polynomials, i.e., the intensity levels of the laminar phases.
While the number of unstable directions unboundedly grows with $\Theta$ for conservative delay, the limit map dynamics, where we have $\Theta=\infty$ indicates that the number of unstable directions has a finite limiting value for dissipative delay.
This means that for dissipative delays and large enough $\Theta$, the attractor dimension is small compared to the values observed for conservative delay.
This picture is confirmed by Fig.~\ref{fig:arnoldkydim}:
To investigate the connection between the dynamics of the access map, Eq.~\eqref{eq:access_map}, and the Kaplan-Yorke dimension of Eq.~\eqref{eq:sys}, in the top panels a heat map of the Lyapunov exponent $\lambda[R]$ (Lyapunov graph \cite{de_figueiredo_lyapunov_1998}) is shown for a periodic delay in (a) and for a quasiperiodic delay with $N=2$ frequencies in (b), where the delay amplitude $A$ and the mean delay $\tau_0$ are varied.
For the further analysis we set the amplitude to $A=0.9$ (white dashed line in the top panels) and vary only the mean delay $\tau_0$.
The Lyapunov exponent $\lambda[R]$ of the access map as a function of $\tau_0$ is shown in the center panels, where the condition for laminar chaos is fulfilled below the black dashed line. 
We then computed the Kaplan-Yorke dimension $D_{KY}$ of Eq.~\eqref{eq:sys} with $f(z)=3.8\,z(1-z)$ for these delay parameters and the result is shown in the bottom panels (blue solid lines)
\footnote{The Kaplan-Yorke dimension was computed adapting the method from \cite{farmer_chaotic_1982}, where the linearized DDE was discretized with step size $\tau_{\text{max}}/M$ with $\tau_{\text{max}}=\max_t \tau(t)$ and $M=2000$.}.
For conservative delays, where we have $\lambda[R]=0$, high-dimensional turbulent chaos is observed, which leads to the local maxima of the Kaplan-Yorke dimension $D_{KY}$.
The local maxima are roughly proportional to $\tau_0$ as we expect due to the equivalence of a system with conservative delay to a non-autonomous system with constant delay (see Appendix~\ref{sec:app_trafo} and \ref{sec:app_KYdim_cons}).
In contrast, the Kaplan-Yorke dimension is drastically reduced for dissipative delay, where we have $\lambda[R]<0$.
Inside each connected region with $\lambda[R]<0$, we observe that the smaller the access map Lyapunov exponent $\lambda[R]$, the smaller is the attractor dimension, which nicely confirms the prediction from the theory of generalized laminar chaos, where, according to Eq.~\eqref{eq:genlamchaos_crit}, a smaller value of $\lambda[R]$ leads to a smaller number of unstable directions of the limit map, Eq.~\eqref{eq:limit_map}.
Considering the whole parameter space, this reduction of the effective dimensionality of the dynamics depends in a fractal manner on the delay parameters $A$ and $\tau_0$, which reflects the fractal structure of the parameter space of the access map.
For periodic delay, $N=1$, the dynamics of the access map can be mapped to the circle map, Eq.~\eqref{eq:circle_map}.
Dissipative delays are associated to parameter regions that lead to mode locking dynamics of the access map with $\lambda[R]<0$.
These regions are the so-called Arnold tongues.
Each tongue is related to a specific rotation number, which is a rational multiple of the map period $\nu_1^{-1}$ and equals the origin of the tongue at $A=0$ \cite{arnold_small_1961,*arnold_small_1961_erratum}.
The remaining set, which is associated to conservative delays, leads to marginally stable quasiperiodic circle map dynamics, $\lambda[R]=0$, and is, for each fixed $A$, a Cantor set with a nonzero measure, where the measure decreases with increasing $A$ and eventually vanishes for $A=1$ \cite{ott_chaos_2002}.
For quasiperiodic delay, the dynamics of the access map, Eq.~\eqref{eq:access_map} can be mapped to the torus map, Eqs.~(\ref{eq:reduced_access_map},\ref{eq:reduced_access_map_component}).
The structure of the parameter space of this map is very similar to the structure of the circle map parameter space with the difference that the generalized Arnold tongues are associated to resonances with multiple frequencies (see Eq.~\eqref{eq:roundtrip_time} and text above) instead of a resonance with only one frequency as in the case of the circle map \cite{petrov_torus_2003}.
As a consequence, for quasiperiodic delay the structures in the parameter space of the access map repeat quasiperiodically in $\tau_0$, which leads to a quasiperiodic modulation of the Kaplan-Yorke dimension $D_{KY}$ with respect to $\tau_0$.
This is in contrast to the periodic modulation of $D_{KY}$ observed for periodic delay, which is caused by the periodic structure of the parameter space with respect to $\tau_0$.
The latter follows directly from the periodicity of the circle map, Eq.~\eqref{eq:circle_map} with respect to $\tau_0$.
 
To investigate the effect of a larger number of frequencies, we performed the same analysis for a quasiperiodic delay with $N=20$ frequencies.
The results are presented in Fig.~\ref{fig:arnoldkydim}(c).
We have chosen the frequencies $\nu_1,\nu_2,\dots,\nu_N$ randomly from a Gaussian distribution with zero mean and standard deviation $\sigma=2$ such that the derivative $\dot{R}(t)=1-\dot{\tau}(t)$ of the access map is an approximation to smoothed Gaussian white noise with mean $1$ and autocovariance function $C(\Delta t) = [A^2/(2 N)]\, \exp[-(\Delta t)^2 \, 2 \pi^2 \sigma^2 ]$.
This relates our results on systems with quasiperiodic delay to the much more complicated problem of systems with random delay such that we are able to give an outlook on future research in this direction.
The torus map that is related to a quasiperiodic delay with many randomly chosen frequencies is related to circle maps with quenched disorder, which are considered in \cite{muller-bender_suppression_2022}, where the spatial structure of the circle map is parameterized by an certain number of disorder parameters, which are randomly chosen but kept fixed during the iteration of the map.
Roughly speaking, there it is demonstrated that, given a large number of disorder parameters, quasiperiodic dynamics, $\lambda[R]=0$, is suppressed in a large part of the parameter space such that, in the limit of an infinite number of disorder parameters, only mode locking dynamics, $\lambda[R]<0$, is observed.
If the randomly chosen frequencies are interpreted as disorder parameters, one may expect that the Lyapunov exponent $\lambda[R]$ of the access map is negative in a large part of the parameter space if the number of frequencies is large.
This conjecture is confirmed by Fig.~\ref{fig:arnoldkydim}(c).
In the upper panel the parameter space right from the tongue-like structure appears to be relatively homogeneous in comparison with the results for periodic and quasiperiodic delay with a small number of frequencies, which are shown in (a) and (b), respectively, while $\lambda[R]$ seems to be always negative at least for larger amplitudes $A$.
This is confirmed by the results in the center panel, where $\lambda[R]$ is plotted as a function of $\tau_0$ for a fixed amplitude $A=0.9$ (blue solid line).
As a consequence the measure of parameter sets with $\lambda[R]=0$, which lead to turbulent chaos is so small that we seem to observe generalized laminar chaos over the whole considered $\tau_0$ range.
The Kaplan-Yorke dimension $D_{KY}$ (blue solid line) for $A=0.9$ appears to be reduced in the whole range in comparison to a system with a constant delay, $A=0$ (black solid line), where turbulent chaos is observed.
For this comparison system, we have set $\tau_0=-v[R]$, where $v[R]$ is the drift velocity, Eq.~\eqref{eq:drift_velocity} of the access map of the time-varying delay system.
As argued in Appendix~\ref{sec:app_KYdim_cons}, one can expect that the local maxima of the Kaplan-Yorke dimension of the time-varying delay is close to the Kaplan-Yorke dimension of the comparison system, which is confirmed in the bottom panels of Figs.~\ref{fig:arnoldkydim}(a) and (b).
Equation~\eqref{eq:sys} with a conservative delay can be transformed to Eq.~\eqref{eq:systrans}, with a time-varying coefficient, which causes the deviations in the Kaplan-Yorke dimension from the autonomous comparison system as argued in Appendix~\ref{sec:app_KYdim_cons}.
For $N\to\infty$, the time-varying coefficient becomes constant so that for increasing $N$ we would expect that the deviations decrease, which is confirmed if Fig.~\ref{fig:arnoldkydim}(a) is compared with (b) but it is contradicted by Fig.~\ref{fig:arnoldkydim}(c), where the deviations are larger compared to (a) and (b). 
These results indicate that for a large number of randomly chosen frequencies, high-dimensional turbulent chaos becomes a very rare phenomenon while generalized laminar chaos dominates, which leads to the overall reduction of the effective dimensionality compared to the autonomous comparison system.
Since the limit of an infinite number of frequencies, $N\to\infty$, leads to a random delay variation, these result can be generalized to systems with random delay, which, however, requires a detailed analysis of the influence of disorder on time-delay systems and will be done elsewhere \cite{muller-bender_laminar_2022}.

\section{Summary}

In this paper we have demonstrated that laminar chaos, a type of chaos known from systems with periodically time-varying delays \cite{muller_laminar_2018,muller-bender_resonant_2019}, can also be observed in systems with quasiperiodic delay.
Therefore we generalized the theory of so-called conservative and dissipative delays introduced in \cite{otto_universal_2017,muller_dynamical_2017} for systems with periodically time-varying delay.
We demonstrated that also quasiperiodically time-varying delays can be classified by the dynamics of the so-called access map, a one-dimensional map that is defined by the time-varying delay and that models the roundtrip dynamics of signals inside a delayed feedback loop, where each delay class leads to different types of chaotic dynamics.
While for a periodic delay the dynamics of the access map can be described by a (one-dimensional) circle map, for a quasiperiodic delay, the dynamics of a torus map, whose dimension equals the number $N$ of incommensurate frequencies of the delay function, has to be considered.
Conservative delays are associated with marginally stable quasiperiodic dynamics, where the Lyapunov exponent of the access map vanishes, $\lambda[R]=0$, and can be transformed to constant delay systems.
They lead to high-dimensional turbulent chaos, which is known from systems with constant delay \cite{farmer_chaotic_1982,ikeda_high-dimensional_1987} and is characterized by chaotic high-frequent oscillations.
Dissipative delays, however, are associated to stable dynamics of the access map, where the Lyapunov exponent of the access map is negative, $\lambda[R]<0$.
While for periodically time-varying delay classical mode locking known from circle map dynamics (cf. \cite{arnold_small_1961,arnold_small_1961_erratum}) is observed, for quasiperiodically time-varying delay the access map shows generalized mode locking known from the torus maps analyzed in \cite{petrov_torus_2003}.
Generalizing the results for periodic delay \cite{muller_laminar_2018,muller-bender_resonant_2019}, dissipative quasiperiodic delays lead to new types of chaos, which are low-dimensional compared to turbulent chaos.
If the criterion $\lambda[f]+\lambda[R]<0$ is fulfilled, where $\lambda[f]$ is the Lyapunov exponent of the one-dimensional map defined by the nonlinearity of the feedback loop, laminar chaos can be observed.
It is characterized by nearly constant laminar phases, whose intensity levels vary chaotically from phase to phase and follow the dynamics of the one-dimensional map defined by the feedback nonlinearity.
While for periodic delay the durations of the laminar phases vary periodically, we demonstrated that they vary quasiperiodically for quasiperiodic delay following the quasiperiodic dynamics on the unstable manifold of the torus map that is associated to the access map.
We further analyzed the dimension of chaotic attractors of systems with quasiperiodic delay generalizing the results from \cite{muller_laminar_2018} for periodic delay, where we computed the Kaplan-Yorke dimension as a function of the mean delay, while keeping other parameters fixed.
While for conservative delays, where turbulent chaos is observed, the Kaplan-Yorke dimension is large and grows proportionally to the mean delay, as known from constant delay systems \cite{farmer_chaotic_1982}, dissipative delays lead to a drastic reduction of the Kaplan-Yorke dimension, especially if the condition for laminar chaos is fulfilled.
As a consequence of the quasiperiodic structure of the parameter space of the torus map that is associated to quasiperiodic delays, the Kaplan-Yorke dimension is quasiperiodically modulated with respect to the mean delay, which is in contrast to the periodic modulation observed for periodic delays.
For a larger number of frequencies $N$ of the quasiperiodic delay modulation, we found that parameters that lead to conservative delays become rare such that for the discretely sampled range of the mean delay all delays appeared to be dissipative.
As a consequence, we observed an overall reduction of the Kaplan-Yorke dimension compared to a constant delay system, where the delay modulation is switched off while the drift velocity of the access map is kept.
These results may have strong implications for systems with randomly time-varying delay, which can be interpreted as the limit of a system with a quasiperiodic delay, where the number $N$ of frequencies that characterize the quasiperiodic delay modulation is sent to infinity.
This problem will be discussed in a forthcoming publication \cite{muller-bender_laminar_2022}.

It would also be interesting to investigate whether laminar chaos or similar dynamics can be observed in systems with equations of motion different from Eq.~\eqref{eq:sys}.
First results in this direction can be found in \cite{khatun_synchronization_2022,kulminskiy_laminar_2022}.
In the following, we shortly discuss two straight-forward generalizations by which new aspects arising in nonscalar systems and in systems with distributed delays can be separately analyzed.
The details will be elaborated in future publications.
A first generalization is obtained by substituting the scalar variable $z(t)$ by a $d$-dimensional vector $\vec{z}(t)$ and the scalar delayed feedback function $f(z)$ by a $d$-dimensional vector field $\vec{f}(\vec{z})$ while considering one discrete time-varying delay $\tau(t)$ and preserving the structure of Eq.~\eqref{eq:sys}.
First numerical results, where $\vec{f}$ was set to the Hénon map (cf. \cite{henon_two-dimensional_1976}), indicate that laminar chaos can be observed in the resulting system, where the $d$-dimensional map $\vec{z}'=\vec{f}(\vec{z})$ governs the dynamics of the laminar phases.
A second generalization is given by introducing a finite-width distribution of the mean delay $\tau_0$ by replacing $z(R(t))$ in Eq.~\eqref{eq:sys} by $\int d\epsilon\, k(\epsilon)\, z(R(t)-\epsilon)$, where the distribution $k(\epsilon)$ has zero mean and a constant width $\sigma_{\tau}$.
The resulting delayed feedback term can be expressed as $f\circ(k\,*\,z)\circ R(t)$, where $\circ$ and $*$ denote composition and convolution, respectively.
The original theory of laminar chaos holds for $\sigma_{\tau} = 0$, where $k(\epsilon)$ is the delta distribution at $\epsilon = 0$.
For nonzero $\sigma_{\tau}$, the distributed delay can be interpreted as an additional smoothing step inside the feedback loop, which is inserted between the frequency modulation of the signal by the time-varying delay $\tau(t)$ and the amplitude modulation by the nonlinearity $f$.
So the influence is similar to the smoothing that is present for finite $\Theta$ even in Eq.~\eqref{eq:sys} due to the low-pass filter defined by the left hand side.
As a consequence, the laminar phases persist in a modified form if $\sigma_{\tau}$ is sufficiently small such that the access map $t'=R(t)-\epsilon$ shows the same mode-locking behavior for all $\epsilon$ with $k(\epsilon)>0$.
Typically such a nonzero $\sigma_{\tau}$ exists since the mode-locking dynamics is structurally stable (cf. \cite{katok_introduction_1997,he_resonances_2022}).

\begin{acknowledgments}
	The authors gratefully acknowledge funding by the Deutsche Forschungsgemeinschaft (DFG, German Research Foundation) - 456546951.
\end{acknowledgments}

\appendix

\section{Transformation of a conservative quasiperiodic delay to a constant delay}
\label{sec:app_trafo}

In the literature, transformations of systems with time-varying delay to constant delay systems were applied, for instance, for solving DDEs \cite{bellman_computational_1965,brunner_time_2009}, for analyzing mathematical properties of solutions \cite{mallet-paret_analyticity_2014,he_construction_2016}, as well as in engineering and biology \cite{otto_transformations_2017}.
Here, we demonstrate that systems with conservative quasiperiodic delay can be transformed to systems with constant delay, adapting the theory for periodic delays in \cite{otto_universal_2017,muller_dynamical_2017}.
We consider general time-delay systems given by the DDE
\begin{equation}
	\label{eq:sysgen}
	\dot{\vec{x}}(t') = \vec{f}(\vec{x}(t),\vec{x}\bm{(}R(t)\bm{)},t),
\end{equation}
with $R(t) = t-\tau(t)$.
By using a nonlinear timescale transformation and $s=\Phi(t)$ and introducing the new variable $\vec{y}(s)$ given by
\begin{equation}
	\vec{y}(s) = (\vec{x}\circ\Phi^{-1})(s) = \vec{x}(\Phi^{-1}(s))
\end{equation}
Eq.~\eqref{eq:sysgen} can be transformed to an equivalent system
\begin{equation}
	\label{eq:systrafo}
	\vec{y}'(s) = (\Phi^{-1})'(s)\, \vec{f}(\vec{y}(s),\vec{y}(R_c(s)),\Phi^{-1}(s)),
\end{equation}
with the new delay $\tau_c(s)=s-R_c(s)$, which is time-varying in general, where $R_c(s)$ is given by
\begin{equation}
	\label{eq:conju}
	R_c(s) = (\Phi \circ R \circ \Phi^{-1})(s).
\end{equation}
This equation defines a topological conjugacy (cf. \cite{katok_introduction_1997,ott_chaos_2002}) between the access maps
\begin{eqnarray}
	\label{eq:access_map_trafo}
	t_k &=& R(t_{k-1}) = R^k(t_0) \\
	\label{eq:transformed_access_map}
	s_k &=& R_c(s_{k-1}) = R_c^k(s_0),
\end{eqnarray}
with $s_k = \Phi(t_k)$, which means that these two maps show the same kind of dynamics.
A transformation to constant delay is possible if there is a $\Phi(s)$ such that $R_c(s)=s-c$, where $c$ is constant.
For periodic delay, this means that the circle map, Eq.~\eqref{eq:circle_map}, that is associated with the access map, Eq.~\eqref{eq:access_map_trafo} is topological conjugate to a pure rotation \cite{katok_introduction_1997}.
In this case the derivative of the transformation function is equal to the invariant density of the circle map \cite{ott_chaos_2002}, i.e., $\dot{\Phi}(t)$ is invariant under the action of the Frobenius-Perron operator of the access map, $\dot{R}^{-1}(t)\,\dot{\Phi}(R^{-1}(t))=\dot{\Phi}(t)$, which can be derived directly from Eq.~\eqref{eq:conju} with $R_c(s)=s-c$ by a few elementary calculations.
The transformation function itself can be computed by \cite{herman_sur_1979}
\begin{equation}
	\label{eq:transformation_function}
	\Phi(t) = \lim_{K\to\infty} \frac{1}{K} \sum_{k=0}^{K-1} R^k(t) - k\,v[R].
\end{equation}

\begin{figure}
	\includegraphics[width=0.48\textwidth]{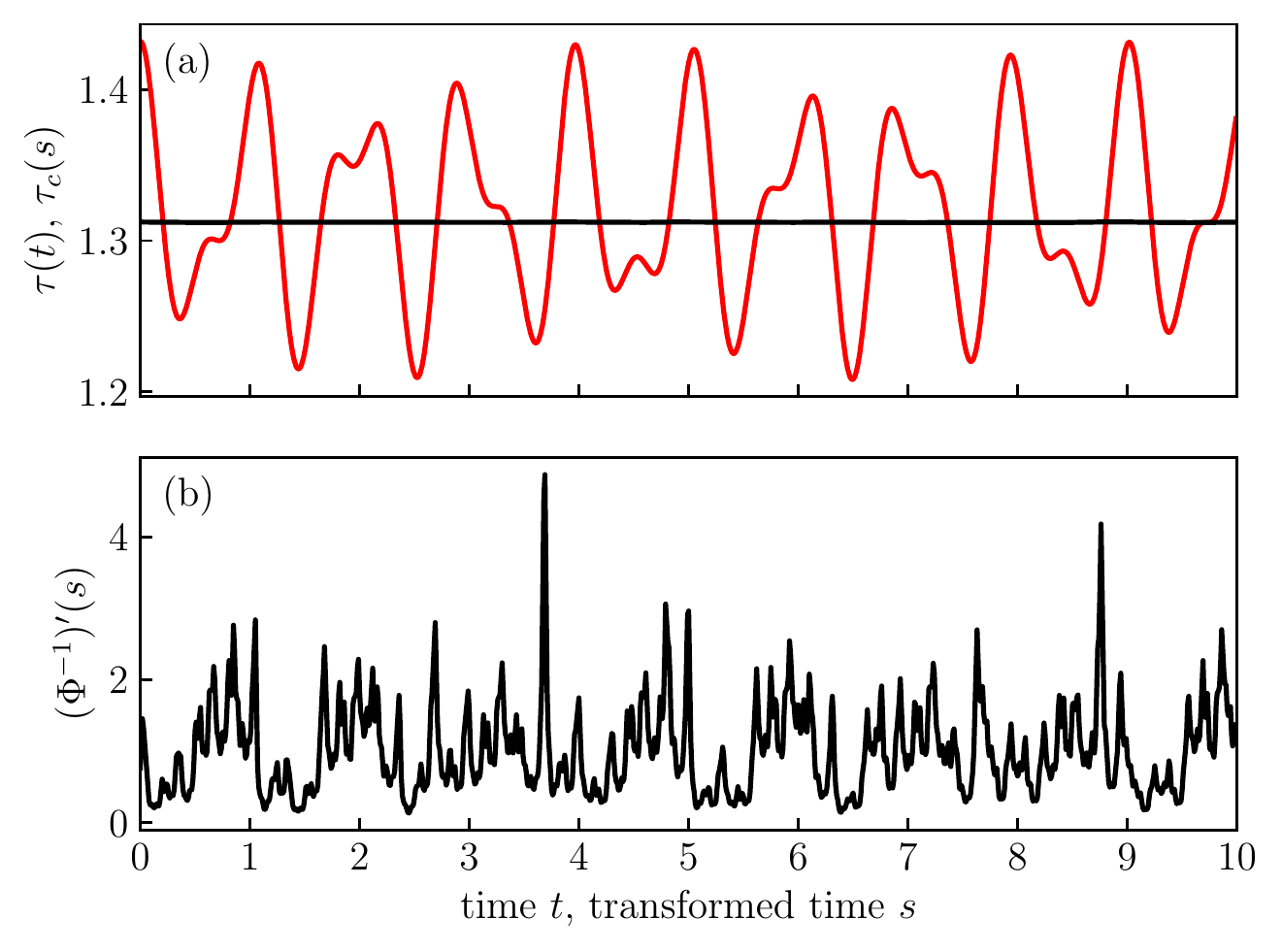}
	\caption{Time scale transformation to constant delay.
		The red line in (a) represents the quasiperiodically time-varying delay $\tau(t)=t-R(t)$ given by Eq.~\eqref{eq:delay_def} with $N=2$, $A=0.9$, $\tau_0=1.32$ that leads to turbulent chaos as shown in Fig.~\ref{fig:turbchaos}(b).
		Using the transformation function given by Eq.~\eqref{eq:transformation_function}, via Eq.~\eqref{eq:conju} $\tau(t)$ can be transformed to the constant delay $\tau_c(s)=s-R_c(s)=-v[R]$ represented by the black line in (a).
		For the numerical computation, we have chosen $K=1000$.
		In (b), the quasiperiodically time-varying coefficient of the equivalent system with constant delay, Eq.~\eqref{eq:systrafo} is shown.
	}
	\label{fig:trafo}
\end{figure}

Since the structure of Eq.~\eqref{eq:conju} does not change when the delay is replaced by a quasiperiodic delay, Eq.~\eqref{eq:transformation_function} can be applied also in this case as demonstrated in Fig.~\ref{fig:trafo}(a), where the conservative quasiperiodic delay used in Fig.~\ref{fig:turbchaos}(b) is transformed to a constant delay.
The resulting quasiperiodically varying coefficient of the transformed system, Eq.~\eqref{eq:systrafo} is shown in Fig.~\ref{fig:trafo}(b).
Applying Eq.~\eqref{eq:transformation_function} to a dissipative quasiperiodic delay would lead to a piecewise constant function, where discontinuities occur at times $t$ whose projections to the phase space of the torus map using Eq.~\eqref{eq:torus_embedding} intersect the unstable manifold of the torus map (see Fig.~\ref{fig:access_map}(c)).
Since in this case $\Phi(t)$ is not invertible, Eq.~\eqref{eq:conju} with $R_c(s)=s-c$ is not fulfilled.
For a conservative quasiperiodic delay, where Eq.~\eqref{eq:transformation_function} converges to a solution of the conjugacy equation, the torus map, Eqs.~(\ref{eq:reduced_access_map},\ref{eq:reduced_access_map_component}), is topological conjugate to the translation on the torus
\begin{equation}
	r_{c,n}(\varsigma) = \varsigma - c\,\nu_n \mod 1.
\end{equation}
Mathematical details on this conjugacy can be found in \cite{petrov_torus_2003}.

\section{Estimate for the Kaplan-Yorke dimension of systems with conservative delay}
\label{sec:app_KYdim_cons}

\begin{figure}
	\includegraphics[width=0.48\textwidth]{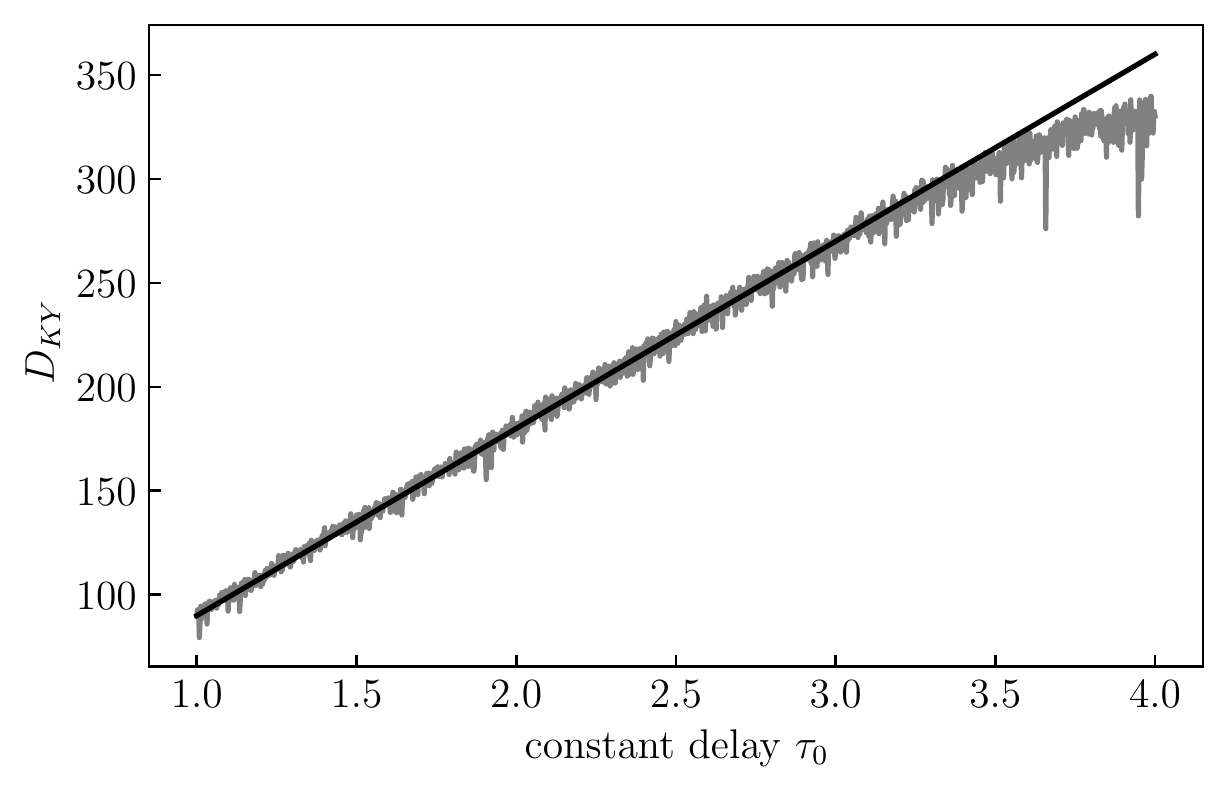}
	\caption{Estimation of the proportional increase of the Kaplan-Yorke dimension according to \cite{farmer_chaotic_1982}. 
		Kaplan-Yorke dimension $D_{KY}$ of Eq.~\eqref{eq:sys} with $\Theta=200$ and $f(z)=3.8\,z(1-z)$ as a function of the constant delay $\tau=\tau_0$ (gray) and the linear fit $D_{KY}(\tau_0) = 90\,\tau_0$ (black).
		The deviations from the linear behavior for large $\tau_0$ are due to the approximation of the infinite dimensional state by a finite number of points and they will vanish for finer discretization.
	}
	\label{fig:kydimconst}
\end{figure}

In this section we analyze the linear scaling behavior of the Kaplan-Yorke dimension $D_{KY}$ of Eq.~\eqref{eq:sys} with a constant delay $\tau=\tau_0$ and argue that it can be used as an upper bound of $D_{KY}$ for the same system with a conservative time-varying delay $\tau=\tau(t)$.
In Fig.~\ref{fig:kydimconst}, the Kaplan-Yorke dimension of this system is shown as a function of the constant delay $\tau_0$ (gray), where the nonlinearity and the parameter $\Theta$ are chosen as in Fig.~\ref{fig:arnoldkydim}.
Except for the largest values of $\tau_0$, where numerical errors become relevant, the Kaplan-Yorke dimension increases linearly (fit in black) in agreement to the early results in \cite{farmer_chaotic_1982}, which were generalized to a large class of constant delay systems in \cite{yanchuk_spatio-temporal_2017}.

Given that the delay is conservative, Eq.~\eqref{eq:sys} can be transformed into a DDE with constant delay and time-varying coefficients, which is according to Eq.~\eqref{eq:systrafo} given by
\begin{equation}
	\label{eq:systrans}
	y'(s) = (\Phi^{-1})'(s)\,\Theta\, \{ -y(s) + f\bm{(}y(s-c)\bm{)} \},
\end{equation}
where $c=-v[R]$ and $v[R]$ is the drift velocity of the access map given by Eq.~\eqref{eq:drift_velocity}.
Given the Lyapunov exponents $\lambda_m$, $m=0,1,\dots$, the Kaplan-Yorke dimension $D_{KY}$ is defined by \cite{kaplan_chaotic_1979,farmer_dimension_1983}
\begin{equation}
	D_{KY} = j + \frac{\sum_{m=0}^{j-1} \lambda_m}{|\lambda_j|},
\end{equation}
where $j$ is the largest integer such that $\sum_{m=0}^{j-1} \lambda_m \geq 0$.
The Lyapunov exponents $\lambda_m$ are the average growth rates of the linearized system 
\begin{equation}
	\label{eq:systranslin}
	(\delta y)'(s) = \alpha(s)\, \delta y(s) + \beta(s)\, \delta y(s-c),
\end{equation}
where $\alpha(s) = -(\Phi^{-1})'(s)\,\Theta$ and $\beta(s) = (\Phi^{-1})'(s)\,\Theta\,f'\bm{(}y(s-c)\bm{)}$ are the partial derivatives of the right hand side of Eq.~\eqref{eq:systrans} with respect to $y(s)$ and $y(s-c)$, respectively.
Equation~\eqref{eq:systrans} and \eqref{eq:systranslin} differ from the autonomous comparison system used in Sec.~\ref{sec:kydim} by the time-varying factor $(\Phi^{-1})'(s)$.

In the following we use the approach from \cite{farmer_chaotic_1982} to demonstrate that this additional factor approximately shifts the Lyapunov spectrum in the negative direction so that the Kaplan-Yorke dimension of this system is smaller than the Kaplan-Yorke dimension of the comparison system.
Then from Eq.~\eqref{eq:transformation_function} and Eq.~\eqref{eq:delay_def} it follows that the factor converges to one in the limit $N\to\infty$ so that the autonomous comparison system gives a good approximation of $D_{KY}$ for systems with conservative delays defined by Eq.~\eqref{eq:delay_def} with large $N$.
First we approximate the tangent space dynamics given by Eq.~\eqref{eq:systranslin} using the Euler method with step size $\Delta s = c/M$.
One step can be compactly written as
\begin{equation}
	\vec{\delta y}_{k+1} = \vec{M}(s_k) \cdot \vec{\delta y}_k,
\end{equation}
where $\vec{\delta y}_k = (s_k, s_{k-1}, \dots, s_{k-M})^{\mathrm{T}}$, with $s_{k+1}-s_k = \Delta s_k$, is the discretization of the memory of the time-delay system at time $s=s_k$ and $\vec{M}(s_k)$ is the $(M+1) \times (M+1)$ matrix
\begin{equation}
	\vec{M}(s) =
	\begin{pmatrix}
		1+\alpha(s)\,\Delta s &  &  &  & \beta(s) \,\Delta s \\
		1 &  &  &  &  \\
		 & 1 &  &  &  \\
		 &  & \ddots\\
		 &  &  & 1 & 
	\end{pmatrix}.
\end{equation}
The long time evolution of an initial volume $V_0$ is described by the sum of the $(M+1)$ Lyapunov exponents of this discrete system.
Assuming that these exponents are good approximations for the largest $(M+1)$ Lyapunov exponents $\lambda_m$ of Eq.~\eqref{eq:systrans}, we obtain the relation between the $\lambda_m$ and the volume $V_K$ after $K$ time steps of the discrete system 
\begin{align}
	e^{K\,\Delta s\, \sum_{m=0}^M \lambda_m} \approx \frac{V_K}{V_0} & = \prod_{k=0}^{K-1} |\det(\vec{M}(s_k))| \\
	& = (\Delta s)^K \prod_{k=0}^{K-1} \beta(s_k) \label{eq:volevo}.
\end{align}
Substituting the approximation
\begin{align}
	\ln\left[ \prod_{k=0}^{K-1} |\beta(s_k)| \right] &= \sum_{k=0}^{K-1} \ln |\beta(s_k)| \\
	& \approx \frac{1}{\Delta s} \int_0^{K\, \Delta s} \ln|\beta(s)|\, ds,
\end{align}
which becomes an identity for $\Delta s \to 0$, into Eq.~\eqref{eq:volevo}, applying the logarithm, dividing by $K\, \Delta s$, computing the difference between the resulting equations for $M$ and $M-1$, and taking the limit $K\to\infty$ gives asymptotically for large $M$
\begin{equation}
	\lambda_M \approx \frac{1}{c} \left[ -\ln M + \ln c + \langle \ln|\beta(s)| \rangle \right],
\end{equation}
where the definition $\langle \,\cdot\, \rangle = \lim_{S\to\infty} \frac{1}{S} \int_0^{S} \cdot \; ds$ is used.
With
\begin{equation}
	\langle \ln|\beta(s)| \rangle =  \langle \ln|\Theta\,f'(y(s-c))| \rangle + \langle \ln|(\Phi^{-1})'(s)| \rangle,
\end{equation}
it follows that the time-varying coefficient $(\Phi^{-1})'(s)$ approximately shifts the Lyapunov spectrum of the autonomous comparison system, which is obtained by setting $(\Phi^{-1})'(s)=1$, by a constant.
Using that the logarithm is a concave function and that we have $\Phi(t) = t + \phi(t)$, where $\phi(t)$ is a quasiperiodic function, we obtain
\begin{equation}
	\langle \ln|(\Phi^{-1})'(s)| \rangle \leq \ln \langle |(\Phi^{-1})'(s)| \rangle = 0.
\end{equation}
where equality holds for the autonomous comparison system, $(\Phi^{-1})'(s)=1$.
So the shift induced by the time-varying coefficient to the Lyapunov spectrum is negative, which leads to a reduction of the Kaplan-Yorke dimension compared to the autonomous comparison system.

\bibliography{ddelamchaosquasi}

\end{document}